\documentclass[prd,aps,twocolumn,a4paper,floatfix,nofootbib]{revtex4}

\usepackage{graphicx,psfrag}
\usepackage{mathrsfs}
\usepackage{amsmath,amsfonts,amssymb}
\usepackage{multirow}
\usepackage{comment}
\usepackage{hyperref}
\usepackage{placeins}
\usepackage[normalem]{ulem} % To get strikethrough (\sout)

\newcommand{\be}{\begin{equation}}
\newcommand{\ee}{\end{equation}}
\newcommand{\bea}{\begin{eqnarray}}
\newcommand{\eea}{\end{eqnarray}}
\newcommand{\bel}{\begin{align}}
\newcommand{\eel}{\end{align}}

\def\GMc2{G M_{\odot} c^{-2}}

\def\E{{E}}

\usepackage{color}

\definecolor{cyan}{rgb}{0,0.9,0.9}
\definecolor{orange}{rgb}{0.9,0.5,0}
\definecolor{magenta}{rgb}{1,0,1}
\definecolor{purple}{rgb}{0.8,0.4,0.8}
\definecolor{gray}{rgb}{0.8242,0.8242,0.8242}

\begin{document}

\title{Mergers of binary neutron stars with realistic spin} 

\author{Sebastiano \surname{Bernuzzi}}
%\email{sebastiano.bernuzzi@uni-jena.de}
\author{Tim \surname{Dietrich}}
%\email{tim.dietrich@uni-jena.de}
\affiliation{Theoretical Physics Institute, University of Jena, 07743 Jena, Germany}
\author{Wolfgang \surname{Tichy}}
%\email{wolf@fau.edu}
\affiliation{Department of Physics, Florida Atlantic University, Boca Raton, FL 33431 USA}
\author{Bernd \surname{Br\"ugmann}}
%\email{bernd.bruegmann@uni-jena.de}
\affiliation{Theoretical Physics Institute, University of Jena, 07743 Jena, Germany\mbox{}}
 
\date{\today}

\begin{abstract} 
Simulations of binary neutron stars have seen great advances in terms
of physical detail and numerical quality. However, the spin of the
neutron stars, one of the simplest global parameters of binaries, 
remains mostly unstudied. 
We present the first, fully nonlinear general relativistic dynamical
evolutions of the last three orbits for constraint satisfying initial
data of spinning neutron star binaries, with astrophysically realistic
spins aligned and antialigned to the orbital angular momentum. 
The initial data are computed with the constant rotational velocity approach.
The dynamics of the systems is analyzed in terms of
gauge-invariant binding energy vs.\ orbital angular momentum
curves. By comparing to a binary black hole configuration we can estimate
the different tidal and spin contributions to the binding energy for
the first time.
First results on the gravitational wave forms are presented. 
The phase evolution during the orbital motion is significantly
affected by spin-orbit interactions, leading to delayed or early
mergers. 
Furthermore, a frequency shift in the main emission mode of the hyper
massive neutron star is observed. 
Our results suggest that a detailed modeling of merger waveforms 
requires the inclusion of spin, even for the moderate magnitudes observed
in binary neutron star systems.
\end{abstract}


\maketitle

\section{Introduction}
\label{sec:intro}

Neutron stars in binaries are spinning objects~\cite{Lorimer:2008se}.
The most famous example is the double pulsar PSR J0737-3039 for which
the orbital period, both spin periods, as well as  
both spin down rates are known~\cite{Lyne:2004cj}. 
The faster spinning pulsar in this system has a spin 
period of $P=22.70$~ms (PSR J0737-3039A)~\cite{Burgay:2003jj}, which 
corresponds to a dimensionless spin of
$\chi\sim0.02$~\cite{Tichy:2011gw,Damour:2012yf,Brown:2012qf,Hannam:2013uu}.
This is the fastest spinning pulsar in a binary system observed so far.
From the orbital period we can estimate that
this system will merge in about $85$~My due to emission of
gravitational waves (GWs).
Over this time period the faster spin will decrease by only about
$20\%$ if we assume spin down is due to magnetic dipole 
radiation~\cite{Tichy:2011gw}. Thus we do expect spin effects
near merger, even though the other spin is much smaller and plays no big
role for this system.

A value of $\chi\sim0.02$ may appear rather small.  For black
hole systems $\chi\leq1$ is expected to approach $1$ in some
cases.
A theoretical limit for isolated neutron stars described by a
large class of nuclear equations of state and uniform 
rotation is $\chi\sim0.7$~\cite{Lo:2010bj}. Configurations with
$\chi>1$ are however possible if, for example,
differential rotation is allowed, e.g.~\cite{Duez:2004nf}. 
It is not clear whether or how many of these
large spin, single neutron stars can be found in binaries.
Theoretical limits for neutron star in binaries depend on the
mechanism of binary formation and binary history 
and are difficult to predict precisely~\cite{Lorimer:2008se}.
Given that the observed neutron star spin in binaries is
comparatively small, star rotation is often ignored when modeling
likely astrophysical neutron star mergers.

Binary neutron stars (BNS) 
are a primary source of GWs. Advanced
interferometric configurations in LIGO and Virgo experiments are
expected to detect from $0.4$ to $400$ events per year, starting from
2018-2019 (or even from
2016)~\cite{Abadie:2010cf,Aasi:2013wya}. At the expected
sensitivities, neglecting spin effects in template-based searches of
BNS can lead to substantial losses in the matched filter signal-to-noise
ratio for the inspiral~\cite{Ajith:2011ec,Brown:2012qf}.   
Template waveforms of the inspiral phase that cover most of the
relevant frequency band are typically constructed with postNewtonian
approximants.  
However, of particular interest is also the detection of the
late-inspiral-merger waveforms, because such signals can be used to
constrain the high-density equation of state (EOS) of neutron
stars~\cite{Read:2009yp,Damour:2012yf,DelPozzo:2013ala}.    
Differently from the inspiral case, precise merger waveforms can be
constructed only by means of numerical relativity
simulations, e.g.~\cite{Baiotti:2011am,Bernuzzi:2011aq,Bernuzzi:2012ci,Hotokezaka:2013mm}.

Although spin is one of the elementary parameters of a binary system,
most studies of BNS to date have not
considered neutron stars with realistic rotation.
Almost all BNS simulations have started
from initial data which have been constructed as (quasiequilibrium)
stationary solutions in circular orbits, within either the
corotational~\cite{Baumgarte:1997xi,Mathews:1997pf,Marronetti:1998xv}
or the irrotational~\cite{Bonazzola:1998yq,Marronetti:1999ya,
  Uryu:1999uu,Taniguchi:2002ns,Uryu:2005vv,Uryu:2009ye} approach.
There is spin in the corotational case, but it is determined by the
orbital period and describes an unrealistic configuration 
because of the low viscosity of neutron star 
matter~\cite{Bildsten:1992ApJ...400..175B}. 

Numerical simulations of BNS mergers in full
general relativity have reached a high degree of precision and
detail. Recent developments include radiation
transport~\cite{Sekiguchi:2011zd}, microphysical equations of 
state~\cite{Sekiguchi:2011mc}, nonideal
magnetohydrodynamics~\cite{Palenzuela:2013hu}, as well as highly  
eccentric
mergers~\cite{Gold:2011df,East:2012ww}. See~\cite{Faber:2012rw} for a
review and more references. In all these simulations the 
Einstein equations are solved without any approximation as a 3+1
evolution system for a given initial configuration. 
Most BNS
simulations have focused on irrotational configurations. 
In this case the stars' spin is neglected, and not modeled in the
simulations. A way to construct quasiequilibrium BNS initial data 
in circular orbits with spins has been recently proposed
in~\cite{Tichy:2011gw,Tichy:2012rp}. The 
\emph{constant rotational velocity} (CRV) formalism developed there
is, to date, the only consistent method to produce realistic initial
data for BNS mergers with spins
(see~\cite{Marronetti:2003gk,Baumgarte:2009fw}
for earlier approximate approaches).

In this work we report dynamical evolutions of BNS
initial data constructed with the CRV approach. 
We study the dynamics of the last three orbits, merger and postmerger
phase, of equal-masses BNS configurations with spins aligned or
antialigned to the orbital angular momentum. The rotational period of
each star is moderate and compatible with astrophysical observations.
We propose two simple ways to estimate the dimensionless spins
of the binary and show that both agree within $\sim10\%$.
The dynamics and gravitational radiation emitted are systematically
compared with an irrotational configuration with the same rest mass.
The orbital evolution is studied by means of gauge-invariant binding
energy vs.\ orbital angular momentum curves~\cite{Damour:2011fu}. We
compare these curves with a binary black hole simulation and with analytical
models, show consistency of the results, and extract the different
contributions to the binding energy from spin and tidal interactions. 
The merger remnant is also investigated, focusing in particular on the
effect of rotation on the hypermassive neutron star.

Our results are the fundamental first step towards the use of 
CRV initial data for modeling rotating stars in BNS mergers. 
In particular, we show that even moderate spins have a significant
impact on the merger dynamics and on the gravitational radiation
emitted. 
Numerical relativity simulations aiming at an accurate description of
the gravitational waves emitted by these sources should take into
account the rotation of the star.

General relativistic evolutions of spinning neutron stars 
have been considered for a long time in the corotational approach, 
both in full general relativity and in the conformally flat approximation,
see e.g.~\cite{Shibata:1999wm,Oechslin:2007gn} 
and~\cite{Faber:2012rw} for other references. 
More recently, alternative approaches have been
proposed in~\cite{Kastaun:2013mv,Tsatsin:2013jca}. Both works employ    
constraint-violating initial data produced by superposing either 
two boosted single-star configurations or an arbitrary velocity
pattern. Such data violate both Einstein constraints and some
hydrodynamical stationarity conditions.
It is unclear how these initial data relate with the ones used in this
work. Thus, in the following, we do not attempt a direct comparison of
the results, but just point out certain similarities. 

The paper is organized as follows. In Sec.~\ref{sec:ID} we review main
aspects of the initial data and describe how to estimate the spin of
our configurations. The numerical method is summarized in
Sec.~\ref{sec:nummeth}. The dynamics of the numerical evolutions is
analyzed in Sec.~\ref{sec:dyn}, by considering: (i)~the analysis of
the orbital motion with binding energy vs.\ orbital angular
momentum curves; and (ii)~the postmerger phase 
and a mode analysis of the hyper-massive neutron star in the merger
remnant. Gravitational radiation is discussed in Sec.~\ref{sec:gws}. We
conclude in Sec.~\ref{sec:conc}. 

Dimensionless units $G=c=M_\odot=1$ are employed hereafter, physical
units are sometimes explicitly given in the text for clarity.

\section{Equilibrium configurations}
\label{sec:ID}

\begin{table*}[t]
  \centering  
  \caption{ \label{tab:models} 
    BNS configurations considered in this work. 
    All initial data are for equal mass configurations,
    where each star has a baryonic mass $M_b=1.625$.
    The polytropic exponent and constant are $\Gamma=2$ and
    $K=123.6489$.
    Spins are aligned or antialigned to the orbital angular
    momentum. 
    The columns contain the following information: 
    the name of the configuration, 
    the rotational part of the fluid velocity given in terms of the
    angular velocity $\omega^z$, 
    ADM mass and ADM angular momentum of the binary,
    the gravitational mass $M_s$ of a single star in isolation, 
    the spin $S_s$ of an isolated star with same $\omega^z$ and $M_b$, 
    and the corresponding dimensionless spin $\chi_s$,
    the spin estimate $S$ using the irrotational configuration as reference point, 
    and the corresponding dimensionless spin $\chi$. 
    $\Gamma$ configurations are evolved with $\Gamma$-law EOS, $P$
    configurations with the polytrope (barotropic evolutions).}
   \begin{tabular}{c|ccccccccc}        
     \hline  
    Name & $\omega^z$& $M_{\rm ADM}$ & $J_{\rm ADM}$ & $M_s$ & $S_s$ &$\chi_s$& $S$& $\chi$ \\
    \hline
    $\Gamma_{050}^{--}$ & -0.00230 & 2.99932 & 8.69761  & 1.51496 &-0.11449 &-0.0499 &-0.10224 &-0.0419 \\
    $\Gamma_{025}^{--}$ & -0.00115 & 2.99911 & 8.79949 & 1.51487 &-0.05710 &-0.0249 &-0.05130 &-0.0198 \\
    $\Gamma_{000}$  &  0.00000 & 2.99903 & 8.90209  & 1.51484 & 0.00000 & 0.0000 & 0.00000 & 0.0000 \\
    $\Gamma_{025}^{++}$ &  0.00115 & 2.99907 & 9.00585 & 1.51487 & 0.05710 & 0.0249 & 0.05188 & 0.0252 \\
    $\Gamma_{050}^{++}$ &  0.00230 & 2.99926 & 9.11092  & 1.51496 & 0.11449 & 0.0499 & 0.10442 & 0.0480 \\
    \hline  
    $P_{100}^{--}$ & -0.00460 & 3.00012 & 8.49472  & 1.51533 &-0.23128 &-0.1007 &-0.20368 &-0.0861 \\
    $P_{000}$  &  0.00000 & 2.99903 & 8.90209  & 1.51484 & 0.00000 & 0.0000 & 0.00000 & 0.0000 \\
    $P_{100}^{++}$ &  0.00460 & 2.99993 & 9.32688 & 1.51533 & 0.23128 & 0.1007 & 0.21240 & 0.0950 \\
    \hline  
   \end{tabular} 
 \end{table*}

\subsection{A review of the CRV approach}
\label{sec:ID:crv}

The initial data used here are constructed using the CRV 
method~\cite{Tichy:2011gw,Tichy:2012rp}. For this method
we use the Wilson-Mathews approach~\cite{Wilson:1995uh,Wilson:1996ty},
which is also known as conformal thin sandwich formalism~\cite{York:1998hy},
for the metric variables together with certain assumptions.
The first assumption is the existence of an approximate helical
Killing vector $\xi^{\mu}$, such that
\be
\pounds_{\xi} g_{\mu\nu} \approx 0 .
\ee
We also assume similar equations for scalar matter quantities such as 
the specific enthalpy $h$. However, the 
4-velocity $u^{\mu}$ is treated differently and it is not assumed that
$\pounds_{\xi} u^{\mu}$  vanishes. Instead
we write 
\be
{u}^{\mu} = \frac{1}{h}(\nabla^{\mu}\phi + w^{\mu}) ,
\label{4-vecolicy}
\ee
where $\nabla^{\mu}\phi$ and $w^{\mu}$
are the irrotational and rotational parts of the fluid velocity.
We then assume that
\be
\label{assumption1}
\gamma_i^{\nu} \pounds_{\xi} \left(\nabla_{\nu}\phi\right) \approx 0 ,
\ee
so the time derivative of the irrotational piece of the
fluid velocity vanishes in corotating coordinates.
We also assume 
\be   
\label{assumption2}
\gamma_i^{\nu} \pounds_{\frac{\nabla \phi}{h u^0}} w_{\nu}
\approx 0 ,
\ee
and
\be   
\label{assumption3}
^{(3)}\!\pounds_{\frac{w}{h u^0}} w_i \approx 0,
\ee
which describe the fact that the rotational piece of the fluid velocity 
is constant along the world line of the star center. These latter two
assumptions lead to the name constant rotational velocity method.

For the data considered here we set
\be
\label{w_choice}
{w}^i = \epsilon^{ijk} \omega^j (x^k - x_{C*}^k) .
\ee
where $x_{C*}^i$ is the center of the star (defined as the point with the
highest rest mass density) and where $\omega^i$ is an arbitrarily chosen
angular velocity vector. In~\cite{Tichy:2012rp} we have verified that this
specific choice leads to only a negligible shear,
so that we can avoid any substantial differential rotation.

This method is implemented in the 
SGRID code~\cite{Tichy:2006qn,Tichy:2009yr,Tichy:2009zr}
which is used to construct the initial data. We then import these data into
the BAM code (see below) 
by spectral interpolation onto BAM's grid points.

\subsection{Selected configurations}
\label{sec:ID:conf}

The initial configurations considered in this work are $\Gamma=2$
polytropes, $p=K\rho^\Gamma$, with $K=123.6489$, 
individual rest mass (or baryonic mass)
$M_b=1.625$ and different rotational states. Table~\ref{tab:models} 
summarizes the main properties of the models.
The rotation state of each star is characterized by its
angular velocity $\omega^i$.
For the simulations described here we have chosen $\omega^i$ to point
along the $z$-direction, with the values given in Tab.~\ref{tab:models}.
If we use $P = 2\pi / \omega^z$ to define a spin period for each star, the
different spinning configurations in Tab.~\ref{tab:models} correspond to
periods of $6.7$, $13.4$ and $26.9$~ms. Notice, however, that these periods
are not exactly the spin periods an observer at infinity would
measure. As we show in Appendix~\ref{app:singlestar} 
the spin periods observed at infinity are about 10\% larger. 
The initial data employed in this work are selected from
equilibrium sequences similar to those computed
in~\cite{Tichy:2012rp}; some details are given in
Appendix~\ref{app:IDseq}. 

The individual isolation masses of the irrotational model are 
$M_{\rm  TOV}=1.51484$, which is equivalent to the ADM
mass of a TOV star with the same rest mass as the binary's individual
$M_b=1.625$. 
All the binary models have about the same proper separation of
$D\approx40.4$ ($59$~km). 
The ADM masses differ by a maximum of $0.04\%$. 
The CRV formalism~\cite{Tichy:2011gw,Tichy:2012rp}
allows us to construct
single rotating star configurations by assuming that the approximate
Killing vector $\xi^{\mu}$ is the timelike Killing 
vector $\xi^{\mu}=(1,0,0,0)$. We have thus computed
single star models with half the rest mass of the binary and the
same $\omega^z$. Each model is characterized by an ADM mass
$M_s$ and an ADM angular momentum $S_s$. For the nonrotating model of
course $M_s=M_{\rm TOV}$ and $S_s=0$; other values are reported in
Tab.~\ref{tab:models}. We will make use of these values in the
following sections. We define $M=2M_{\rm TOV}$, and scale the time in
the plots with this mass.  

Additionally to these BNS configurations, we consider a
nonspinning equal-masses binary black hole (BBH) run. 
The initial configuration is identical to the one in Tab.~III
of~\cite{Walther:2009ng} with an initial separation of $\approx11$ and
an eccentricity of $\approx 0.0002$.

\subsection{Spin estimates}
\label{sec:ID:spin}

In the CRV approach the natural quantity
describing the spinning motion is $\omega^i$, however in the context
of GWs it is convenient to consider ``a spin''. Since the spin
of a single star in a binary is not unambiguously defined 
in general relativity, we propose here two simple different ways of
estimating the spin magnitude $S$. 

A simple method (which to our knowledge is new in this context)
is to consider single stars in isolation with the same
rest mass and the same $\omega^z$, computed as described in
Sec~\ref{sec:ID:conf}. These stars have a well defined angular
momentum $S_s$. 
We then take the spin to be 
\be
\label{eq:spinestimate1}
S \approx S_s \ , 
\ee
and the dimensionless spin to be $\chi_s=S_s/M_s^2$. These values are
reported in Tab.~\ref{tab:models}. 

A second estimate is given by comparing each spinning configuration with
the irrotational one. From Tab.~\ref{tab:models} we observe that the
spin does not contribute significantly to the ADM masses. Assuming
that the differences in the total angular momentum are due to the
spins of the stars, we write
\be
\label{eq:spinestimate2}
S \approx (J_{\rm ADM} - J_{\rm ADM}^{\rm irr})/2 \ .
\ee
Dimensionless spin values are given then by $\chi=S/M_s^2$.
The results are stated in Tab.~\ref{tab:models} and differ from 
the previous estimate by $\sim 10\%$.

A precise value of the dimensionless spin is necessary to construct
binding energy vs.\ orbital angular momentum curves. We will show that
a nontrivial agreement with analytical results can be obtained using
Eq.~\eqref{eq:spinestimate1}.

\section{Numerical method}
\label{sec:nummeth}

\begin{table}[t]
  \centering  
  \caption{ \label{tab:grid} 
    Summary of the grid configurations used for the evolutions,
    see Sec.~\ref{sec:nummeth} for a detailed description.}
  \begin{tabular}{ccccccccccc}        
    \hline
    Name & $L$ & $l^{\rm mv}$ & $n^{\rm mv}$ &
    $h_{L-1}$ & $n$ & $h_0$ \\
   \hline
   L1 & 6 & 2 & 128 & 0.225  & 128 & 7.20    \\   
   L2 & 6 & 2 & 144 & 0.200  & 144 & 6.40    \\   
   M & 6 & 2 & 168  & 0.171  & 168 & 5.49    \\      
   H & 6 & 2 & 192 & 0.150  & 192 & 4.80    \\
    \hline
  \end{tabular}
\end{table}

Simulations are performed with the BAM
code~\cite{Thierfelder:2011yi,Brugmann:2008zz,Bruegmann:2003aw,Bruegmann:1997uc}.
The Einstein equations are written in 3+1 BSSNOK
form~\cite{Nakamura:1987zz,Shibata:1995we,Baumgarte:1998te}.
1+log and gamma-driver conditions are
employed for the evolutions of lapse and shift, 
respectively~\cite{Bona:1994a,Alcubierre:2002kk,vanMeter:2006vi}; 
see~\cite{Thierfelder:2010dv} for a study of gauge condition performance in
handling gravitational collapse. 
General-relativistic hydrodynamics (GRHD) equations are solved in
conservative form by defining Eulerian conservative variables from the
rest-mass density $\rho$, pressure $p$, internal energy $\epsilon$, and
3-velocity, $v^i$.
An equation of state closes the system.
We consider evolutions with a $\Gamma$-law EOS
\be
p =(\Gamma-1) \rho \epsilon ,
\ee
with $\Gamma=2$ for most of the configurations. Some control runs with a
polytropic EOS, thus forcing a barotropic evolution, have also been
performed (see Tab.~\ref{tab:models}.)

The evolution algorithm is based on the method-of-lines with explicit
4th order Runge-Kutta time integrators. Finite differences (4th order
stencils) are employed for the spatial derivatives of the metric. 
GRHD is solved by means of a high-resolution-shock-capturing 
method~\cite{Thierfelder:2011yi} based on primitive reconstruction
and the Local-Lax-Friedrichs (LLF) central scheme for the numerical
fluxes. Primitive reconstruction is performed with the 5th order WENO
scheme of~\cite{Borges20083191}, which has been found to be important
for long term accuracy~\cite{Bernuzzi:2012ci,Bernuzzi:2011aq}. 
The numerical domain is made of a hierarchy of cell-centered nested
Cartesian grids. The hierarchy consists of $L$ levels of refinement labeled
by~$l = 0,...,L-1$. A refinement level $l$ has one or more
Cartesian grids with constant grid spacing $h_l$ and $n$ points per
direction. The  refinement factor is two such that $h_l = h_0/2^l$. The grids
are properly nested in that the coordinate extent of any grid at
level~$l$, $l > 0$, is completely covered by the grids at level~$l-1$.
Some of the mesh refinement levels $l>l^{\rm mv}$ can be dynamically moved and adapted
during the time evolution according to the technique of~``moving boxes''.
The Berger-Oliger algorithm is employed for the time stepping~\cite{Berger:1984zza}, 
though only on the inner levels. Interpolation in 
Berger-Oliger time stepping is performed at second order. A 
Courant-Friedrich-Lewy factor of~$0.25$ is employed in all runs. 
We refer the reader to~\cite{Thierfelder:2011yi,Brugmann:2008zz} for more details.

The grid configurations considered in this work are reported in
Tab.~\ref{tab:grid}. Because we evolve equal-mass binaries, we use
bitant symmetry (evolving only the half space $z>0$) without loss of
generality. We experimentally found that the nonconservative mesh
refinement in BAM can lead to rest mass violations during the postmerger
phase (when mass crosses AMR boundaries), and in turn degrade the
quality of the simulation in the long-term. In order to minimize this
systematic source of error, the number of points in the moving levels
is set equal to the nonmoving ones; see
Appendix~\ref{app:tests} for more details. 
This is different from what was done in previous BAM
simulations, that instead mostly focused on the orbital phase,
e.g.~\cite{Thierfelder:2011yi}. 

Gravitational radiation is computed by means of the Weyl
scalar~\cite{Brugmann:2008zz} on a coordinate sphere of radius $r=400$. The
scalar is projected onto spin weighted spherical harmonics to compute
the multipoles $\psi^4_{lm}$. The metric multipoles $h_{lm}$ are calculated 
by integrating the relation $\psi^4_{lm}=\ddot{h}_{lm}$. We use a
frequency-domain procedure with a low-frequency cutoff~\cite{Reisswig:2011CQGra..28s5015R}. The signal is integrated from the 
very beginning of the simulation, in  order to include also the
initial burst of radiation related to the conformal flatness of the initial
data.
The radiated energy and angular momentum perpendicular to the orbital
plane are calculated as
\begin{align}
\mathcal{E}_{\rm rad} &=\dfrac{1}{16\pi}\sum_{l,m}^{l_{\rm max}} \int_{0}^t dt'\left|r\,\dot{h}_{lm}(t')\right|^2\\
\mathcal{J}_{z\ \rm rad} &=\dfrac{1}{16\pi}\sum_{l,m}^{l_{\rm max}}\int_{0}^{t}d t' m\left[r^2\,h_{lm}(t')\dot{h}_{lm}^*(t')\right] \ ,
\end{align}
with $l_{\rm max}=8$. In the calculation of the total angular momentum
$\mathcal{J}_{\rm rad}$ we also include the $\mathcal{J}_{x,y\ \rm
  rad}$ components, although their contribution is nonzero only in the
postmerger phase and in practice negligible.

\section{Dynamics}
\label{sec:dyn}

In this section we discuss the effect of the star's rotation on the
binary dynamics. We formally define the merger as the peak of the
amplitude $|r\,h_{22}|$ (Sec.~\ref{sec:gws}), but recall that the two
stars come in contact well before (see e.g.~discussion
in~\cite{Bernuzzi:2011aq} and below). 
First we describe the orbital phase, i.e.~evolution up
to merger, then we consider the postmerger phase.

\begin{figure*}[t]
  \centering
    \includegraphics[width=0.32\textwidth]{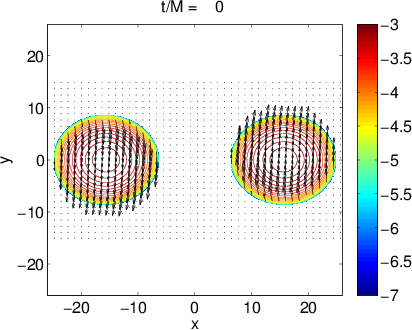} 
    \includegraphics[width=0.32\textwidth]{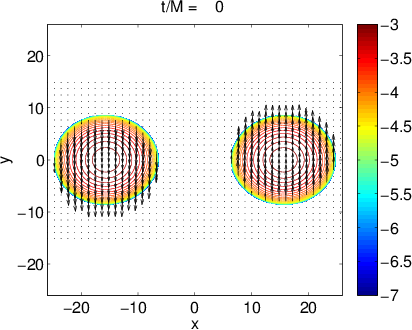} 
    \includegraphics[width=0.32\textwidth]{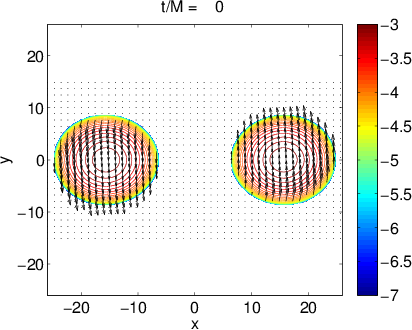}\\    
    \includegraphics[width=0.32\textwidth]{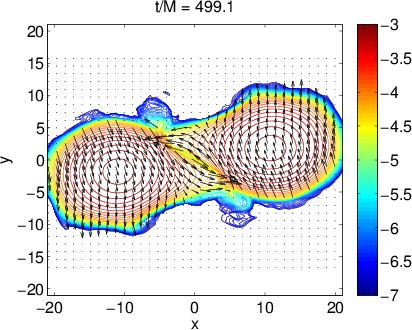} 
    \includegraphics[width=0.32\textwidth]{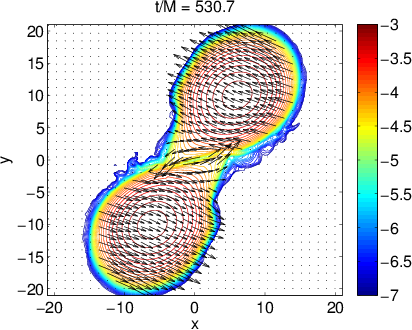} 
    \includegraphics[width=0.32\textwidth]{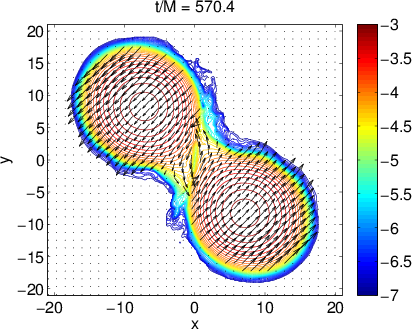}\\
    %`
    \includegraphics[width=0.32\textwidth]{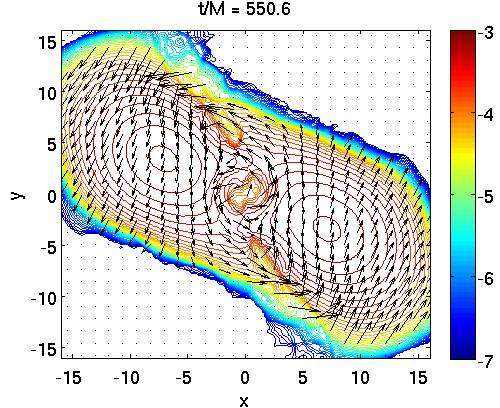}
    \includegraphics[width=0.32\textwidth]{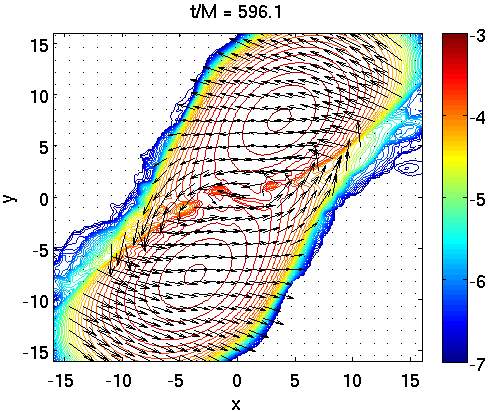}
    \includegraphics[width=0.32\textwidth]{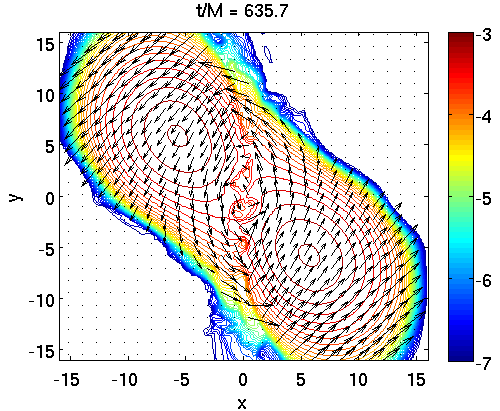}
    \caption{ \label{fig:snap2d} 
      Snapshots of $\log_{10}\rho$ and $(v^x,v^y)$ on the orbital
      plane. Rows from top to bottom refer to initial, contact and 
      merger times.
      Columns from left to right refer to models $\Gamma_{050}^{--}$, $\Gamma_{000}$ and $\Gamma_{050}^{++}$,
      respectively. Note the different spatial scales.}
\end{figure*}

\subsection{Orbital motion}
\label{sec:dyn1}

\begin{figure}[t]
  \centering
    \includegraphics[width=0.5\textwidth]{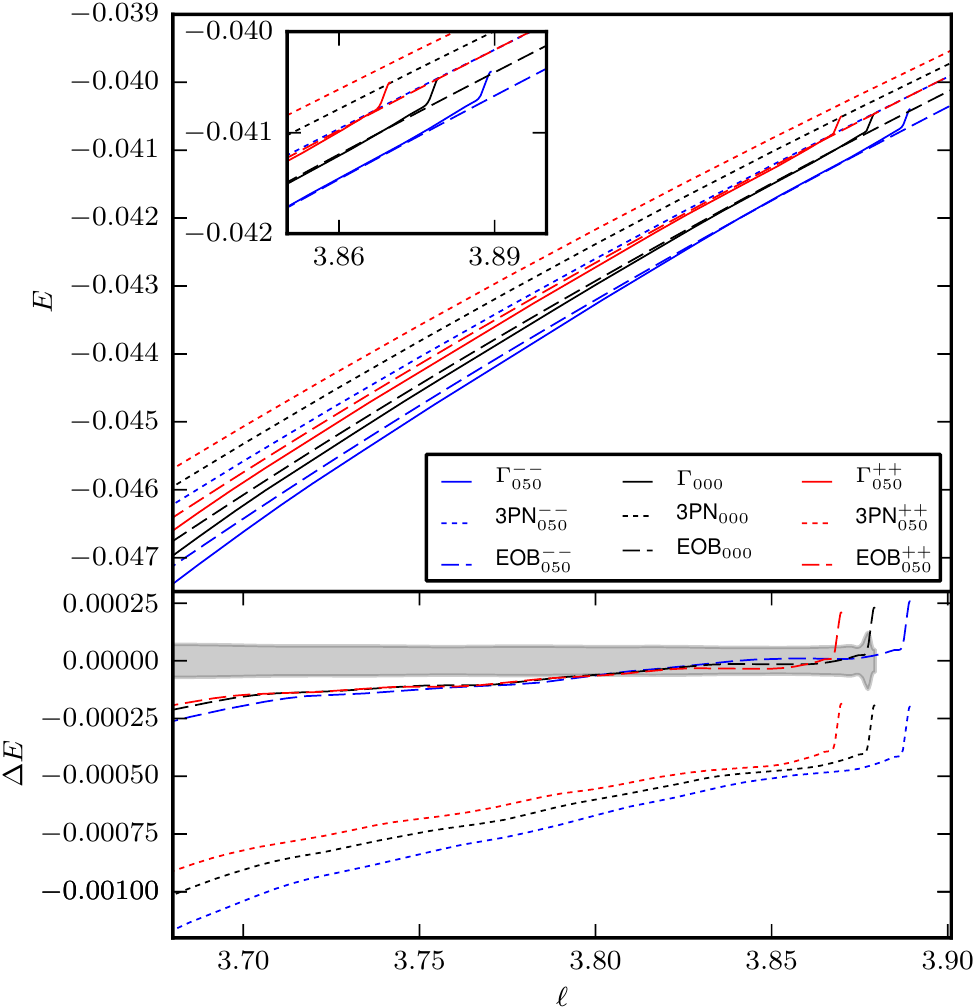}
    \caption{ \label{fig:Ejearly} Binding energy vs.\ orbital angular
      momentum curves for models $\Gamma_{050}^{--}$,
      $\Gamma_{050}^{++}$ and $\Gamma_{000}$.  
      Top: $E(\ell)$ curves for numerical data (solid lines),
      3PN (dotted lines), and EOB (dashed lines).
      Bottom: Differences $\Delta E = E-E^{X}$ between numerical
      data and 3PN (dotted) and EOB curves (dashed). 
      The uncertainty on the numerical data is shown in light
      gray.}
\end{figure}

\begin{figure*}[]
  \centering
  \includegraphics[width=1.\textwidth]{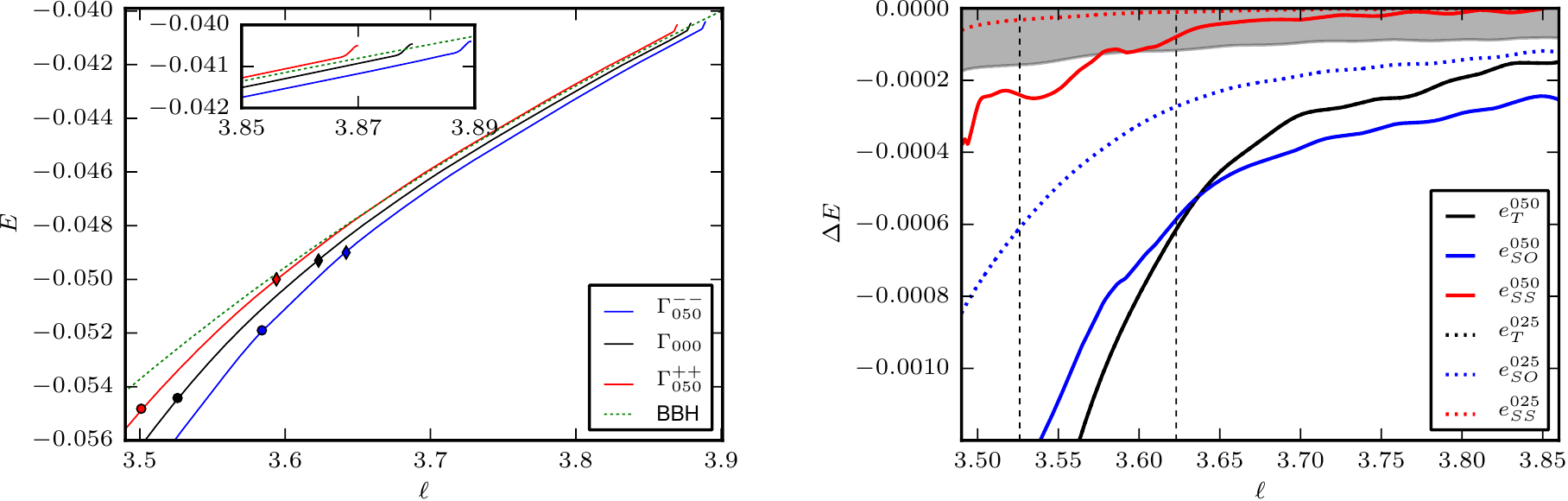}
    \caption{ \label{fig:Ej} 
      Binding energy vs.\ orbital angular
      momentum curves for $\Gamma$-models and a nonspinning BBH run. 
      Left: $E(\ell)$ curves for BNS and BBH data. Diamonds and
      bullets indicate the approximate moment of
      contact (corresponding to the snapshots of
      Fig.~\ref{fig:snap2d}) and the moment of merger, respectively. 
      Right: Different contributions to the binding energy in
      Eq.~\eqref{eq:Epieces}, extracted
      from differences of data sets as described in the text. The
      uncertainty of the numerical data  
      is shown as the gray shaded region. The vertical dashed
      lines refer to the approximate moment of contact and to the merger
      for $\Gamma_{000}$.
      $e^{025}_T$ and $e^{050}_T$ coincide in this plot. 
      Doubling the spin
      approximately doubles the spin-orbit effect in the binding energy.
      For $0.05$ the spin-orbit term is larger than the 
      tidal term until $\ell\sim3.65$.} 
\end{figure*}

Figure~\ref{fig:snap2d} shows snapshots of the rest-mass density and 
fluid's velocity $(v^x,v^y)$ on the orbital plane for the representative
models $\Gamma_{050}^{--}$, $\Gamma_{000}$, and $\Gamma_{050}^{++}$ (columns). We
focus on these since they are the $\Gamma$-law EOS 
evolutions with the highest spin magnitudes. In the plot, 
the stars orbit each other counterclockwise. The top
row refers to the initial time; comparing the central panel
($\Gamma_{000}$) with the left ($\Gamma_{050}^{--}$) and the right 
($\Gamma_{050}^{++}$), one can see only a very small difference in the velocity
pattern due to the rotational state of the CRV data with respect to the
irrotational flow. 
The central row refers to a simulation time at which the cores of the
two stars come in contact, i.e.~rest-mass density layers
$\rho\sim10^{-4}$ ($10^{14}\ \text{g/cm}^3$) of the two stars touch
each other in the characteristic shearing contact,
e.g.~\cite{Sekiguchi:2011zd}.  
The proper distance between the stars, as calculated from the local
minima of the lapse function or local maxima of $\rho$, is about
$D\sim30$ at this moment. 
Note also the very different orbital phases of the three models at
this moment, revealing that the moderate initial spins had a significant
effect after only about $1.5$ orbits. 
The last row refers to the merger time, after approximately three
orbits (six-seven GW cycles, see below), at which hyper massive
neutron stars (HMNSs) are formed for the three configurations.  The
HMNSs appear similar in the snapshots, but their angular momentum is
actually different and different dynamics follows (see Sec.~\ref{sec:dyn2}.)  

The orbital dynamics of the irrotational model is consistent with what was
previously observed in 
e.g.~\cite{Thierfelder:2011yi} for the same initial configuration
computed with the Lorene code (see also Appendix~\ref{app:tests}). 
The star rotation changes this picture: for spins aligned with
the orbital angular momentum the inspiral is longer for larger spin
magnitudes, while for antialigned spins the inspiral is shorter 
for larger spin magnitudes. This effect can be understood in term of
spin-orbit interaction~\cite{Damour:2001tu}. 
Analogously to what happens to corotating/counter-rotating circular
orbits in Kerr spacetimes, the last-stable-spherical orbit moves
outwards (inwards) for antialigned (aligned) spin configurations with 
respect to the nonspinning case~\cite{Damour:2001tu}. 
The analogous result in binary black hole
simulations is sometimes called ``hang-up''~\cite{Campanelli:2006uy}. 
In BNS mergers it has been discussed recently in~\cite{Kastaun:2013mv,Tsatsin:2013jca}.
Spin-orbit interactions thus change quantitatively the binary
dynamics, and we quantify this aspect in the following.

A gauge invariant way to analyze the binary dynamics from numerical
relativity data is to consider binding energy  vs.\ orbital angular
momentum curves, as proposed in~\cite{Damour:2011fu}.  
In the present context these curves allow us to characterize the
dynamics generated by CRV initial data. 
We compute the dimensionless binding energy and angular momentum per
reduced mass as~\footnote{ 
  The notation for the angular momentum is slightly different
  from~\cite{Damour:2011fu}.}
\begin{align}
\label{Ebdef}
\E &= \left[(M_{\rm ADM}(t=0)- \mathcal{E}_{\rm rad})/M-1\right] \nu^{-1} \\
\label{jdef}
\ell  &= (L - \mathcal{J}_{\rm rad})(M^2\nu)^{-1} \ , 
\end{align}
respectively, where $\nu=1/4$ is the symmetric mass ratio, and the
isolation mass $M$ is taken as $M=2 M_s$, see Table~\ref{tab:models}. 
The initial angular momentum $L$ is computed from the
spin estimates of Sec.~\ref{sec:ID:spin} as 
\be
L = J_{\rm ADM}(t=0) - 2 S_s \ ,
\ee
and coincides with $ J_{\rm ADM}(t=0)$ for the irrotational
configuration. Eq.~\eqref{jdef} assumes that $\mathcal{J}_{\rm rad}$
only affects $\ell$, i.e. the spin magnitude remains constant. 
This can be justified on a PN basis, and, in general, it holds 
for small, aligned spins.

\begin{table}[t]
  \centering  
  \caption{ \label{tab:dynamics} 
    Dynamical quantities during orbital motion.
    Simulation time, gravitational wave
    frequency, angular momentum, and energy are reported at the
    moments of contact and merger. Note the contact time is not a well
    defined quantity, and just reported to give a rough
    estimate. Frequencies uncertainties of about $10$\%.}   
  \begin{tabular}{c|cccc|cccc}        
    \hline
    Name & $t_c/M$ & $M\omega_c$ & $\ell_c$ & $E_c\times10^{2}$ & $t_m/M$ & $M \omega_m$ & $\ell_m$ & $E_m\times10^{2}$ \\
    \hline
    $\Gamma_{050}^{--}$ & 499 & 0.067 & 3.64 & -4.89 & 551 & 0.124 & 3.58 & -5.19 \\
    $\Gamma_{025}^{--}$ & 514 & 0.065 & 3.63 & -4.90 & 575 & 0.128 & 3.55 & -5.36\\
    $\Gamma_{000}$      & 531 & 0.069 & 3.62 & -4.92 & 595 & 0.127 & 3.53 & -5.44 \\
    $\Gamma_{025}^{++}$ & 549 & 0.070 & 3.61 & -4.95 & 618 & 0.125 & 3.51 & -5.47\\    
    $\Gamma_{050}^{++}$ & 570 & 0.071 & 3.60 & -4.99 & 636 & 0.123 & 3.50 & -5.48 \\
    \hline
    \end{tabular}
\end{table}

The numerical data $E(\ell)$ are compared to \emph{point-mass}
analytical results: a postNewtonian~(PN) and an
effective-one-body~(EOB)~\cite{Buonanno:1998gg,Buonanno:2000ef} 
curve.  
In this work we employ the 3PN binding energy expression including
next-to-next-to-leading order spin-orbit coupling as given by Eq.~(43)
of~\cite{Nagar:2011fx}, and indicate it as $\E^{\rm 3PN}(\ell)$. 
The result rely on earlier achievements in PN theory, among others
see~\cite{Kidder:1992fr,Kidder:1995zr,Tagoshi:2000zg,Blanchet:2004ek,Faye:2006gx,Damour:2007nc,Steinhoff:2007mb}. 
Additionally, we also consider the curve $E^{\rm EOB}(\ell)$
constructed within the EOB approach in the
adiabatic limit. For simplicity, we use the EOB model for
spinning binaries introduced in~\cite{Damour:2001tu}. Similarly, the
nonspinning part of the model is taken at 3PN
accuracy~\cite{Damour:2000we} only and it is resummed with a $(1,3)$
Pad\'e approximant
(see~\cite{Barausse:2011dq,Damour:2012ky,Bini:2013PhRvD..87l1501B,Pan:2013rra} for 
recent theoretical developments of the EOB model.)  
The next-to-leading-order~\cite{Damour:2008qf} and
next-next-to-leading-order~\cite{Nagar:2011fx} spin-orbit
couplings are included in the Hamiltonian.
We restrict ourselves to the leading order spin-spin term for
simplicity, although the spin-spin interaction is known at
next-to-leading order~\cite{Balmelli:2013PhRvD..87l4036B}.

There is evidence that irrotational conformally flat initial data
sequences are quite close to the 3PN result for a sufficiently large 
binary separation,
e.g.~\cite{Uryu:2009ye} (and also Appendix~\ref{app:IDseq}). However, we
recall that the conformally flat approximation introduces errors 
already at 2PN level~\cite{Damour:2000we}. On the other hand, the
3PN-EOB adiabatic curve has been found to correctly reproduce nonspinning 
numerical relativity data of different mass ratios up to
$\ell\sim3.55$~\cite{Damour:2011fu}. The same reference has shown
that the 3PN-EOB curve is instead remarkably close to numerical data
up to the last stable orbit of the EOB potential ($\ell\sim3.28$), and it is an 
accurate diagnostic of the \emph{conservative dynamics} of the system.
(See~\cite{Bernuzzi:2012ci} for the case of neutron star mergers with
irrotational data.)

The curves $\E(\ell)$ at early simulation times are shown in
Fig.~\ref{fig:Ejearly} for models $\Gamma_{050}^{--}$, $\Gamma_{000}$, and
$\Gamma_{050}^{++}$, together with the PN and EOB curves computed with the
spin values as estimated in Sec.~\ref{sec:ID:spin}.
These curves are quite sensitive to small variations in the values of
the initial masses, angular momentum and spins. For example, they require $M_{\rm
  ADM}$ and $M_s$ accurate up to four digits. The uncertainty on the
numerical data is also shown. It is estimated considering $\Gamma_{000}$
data at different resolutions (grid configurations H and L2) and
including the uncertainty of the initial ADM values as measured from
different SGRID resolutions. Both errors are added in quadrature.
The bottom panel shows the differences $\Delta\E=\E-\E^{X}$ of
numerical data with respect to the $X=$~3PN and the $X=$~EOB curves with the
relative spin values.

We experimentally observe that, for all the configurations considered in
this work, the spin estimate in Eq.~\eqref{eq:spinestimate1} leads
to $\E(\ell)$ curves closer to the PN and EOB ones at early times
than Eq.~\eqref{eq:spinestimate2}. 
Thus, we use in the figure and in the following that estimate.
Note that this choice assumes that the spin is almost constant along
the sequences.   

As shown in Fig.~\ref{fig:Ejearly}, the dynamics starts between the PN
and EOB curves, and rapidly depart from the initial state,
$\ell\sim3.87$ (see inset). This variation is due to the emission of the
artificial gravitational radiation related to the conformally flat assumption of
the CRV data. In complete analogy with the nonspinning binary black 
hole case and irrotational case, the numerical evolution settles very
quickly close to the EOB curve (with the proper
spin)~\cite{Damour:2011fu,Bernuzzi:2012ci}.  
The difference between the EOB curves and the numerical data at early
times is within the error-bars: the tidal contribution cannot be
distinguished with present data (the same happens comparing BNS and BBH,
see below). 

A clear hierarchy among the PN and EOB curves with different spins can
be observed. This effect is due to spin-orbit interactions:
antialigned configuration are more bound and aligned configurations
are less bound than irrotational (cf.~``hang-up''). The numerical
curves consistently respect such hierarchy from early times to merger
(see below).  
During the early-times evolution, the binaries binding energies depart
systematically from the EOB, and 
close to contact ($\ell_c\sim3.63$), the
deviation becomes significant. Note in the bottom panel how the
differences between EOB and numerical data for different spins are
essentially indistinguishable. This fact clearly suggests that the
deviation is due to finite size effects.

The curves $\E(\ell)$ up to merger are shown in
Fig.~\ref{fig:Ej} (left panel) for 
models $\Gamma_{050}^{--}$, $\Gamma_{000}$, and $\Gamma_{050}^{++}$ 
together with the one for the nonspinning BBH
run. At early times (see inset) the BBH system is less bound than the
irrotational configuration, but within the data uncertainty. 
As observed for the EOB curve, for $\ell\to\ell_c$ tidal contributions 
become progressively more important and the systems become more bound 
deviating systematically from the BBH curve. Merger occurs at
$\ell_m\sim3.58$, $3.53$, $3.50$ for $\Gamma_{050}^{--}$,
$\Gamma_{000}$ and $\Gamma_{050}^{++}$,  respectively. At merger, the aligned spin
configurations are more bound than the antialigned one.
See also Tab.~\ref{tab:dynamics} for a
collection of relevant numbers for all the configurations.

In order to gain insight into the role
of spin and tidal interactions during the 
merger phase, we make the assumption that
\be
\label{eq:Epieces}
\E \approx e_0 + e_{SO} + e_{SS} + e_T \ ,
\ee
i.e.~that the binding energy of a spinning BNS configuration can be
approximated by the \emph{sum} of four separate contributions: a 
nonspinning point-mass (black-hole) term $e_0$, a spin-orbit (SO) term
$e_{SO}$, spin-spin (SS) term $e_{SS}$, and a tidal (T) term
$e_T$. The different terms have PN contributions starting from
1.5PN~(SO), 2PN~(SS) and 5PN~(T).
All the four terms in~\eqref{eq:Epieces} can be calculated using the
simulation data, e.g.~the four runs $\Gamma_{000}$, $\Gamma^{++}_{050}$, 
$\Gamma^{--}_{050}$, and BBH. Below we distinguish between 
the terms in the ansatz ($e_x$) and the numerical curves
($\E^Y_X$, as the relative model name).
The SO term has structure of the form 
$\propto\mathbf{L}\cdot\mathbf{S}$, so for aligned/antialigned spins
$e_{SO}\propto2\,{\rm sign}(S)|\mathbf{L}||\mathbf{S}|$.
Similarly, the SS term has structure
$\propto\mathbf{S}_1\cdot\mathbf{S}_2$, so it does not change sign if 
both spins flip.  
A $++$ binary configuration has a repulsive SO contribution
($e_{SO}>0$), whereas a $--$ one with the same spin magnitude has an
attractive SO contribution ($e_{SO}<0$) to the binding energy.
However, as mentioned above, the aligned spin configurations give a
more negative binding energy at merger than the antialigned
configurations (compare with~\cite{Damour:2001tu}.) 

The SO term is calculated by the combination of the
aligned/antialigned spins runs with the same magnitude,
i.e.~$e_{SO}\approx(\E^{++}_{050}-\E^{--}_{050})/2$.   
Obviously we pose $e_0\approx \E_{\rm BBH}$ and $e_0 + e_T
\approx\E_{000}$, and calculate $e_T$ from the difference
$\E_{000}-\E_{\rm BBH}$. 
The SS term is estimated as $e_{SS}\approx
(\E^{++}_{050}+\E^{--}_{050})/2 - \E_{000}$. 
The different terms $e_x$ are reported in the right panel of
Fig.~\ref{fig:Ej}. The SS contribution is the smallest negative, at
the level of the uncertainty of the data. At the moderate spins used here,
SS interactions are essentially not resolved in the simulation. 
On the other hand, the curves $e_{SO}$ and $e_T$ are well 
resolved. We observe that, for $\chi=0.05$, the $e_{SO}$ is the
dominant contribution to the binding energy up to
$\ell\sim3.65$. After this point, $e_T$ becomes dominant. 
This corresponds to intuition since the
dynamics reaches the hydrodynamical regime (see
Fig.~\ref{fig:snap2d}). Towards merger (not visible in the plot) the
differences between the $e_{SO}$ and $e_{T}$ become progressively
larger. 

An independent estimate of $e_{SO}$, $e_T$, and $e_{SS}$ is also given
by using the data of the other two simulations $\Gamma_{025}$,  see
right panel of Fig.~\ref{fig:Ej}. We 
obtain similar results, and in particular the $e_T$ terms exactly
coincide as they should.
There is one important difference though. In the case
$\chi=0.025$, the $e_{T}$ term is the largest negative term already at
early simulation times.  Thus during the last three orbits the binding
energy is ``tidally dominated'' as in the irrotational case. 

We mention that, while the SS term is poorly resolved, its presence is
clearly suggested by looking at the difference $\E_{000}-\E^{++}_{050}$ and
$\E^{--}_{050}-\E_{000}$. The two combinations approximate $e_{SO}\pm
e_{SS}$, with the SS term entering with a different sign. We find
that, as expected, the former is less bound, the latter is more by a
small amount. Similarly, inspection of the quantity
$(\E_{050}^{++}-\E_{050}^{--})/2-\E_{BBH} \approx e_T +e_{SS}$, leads
to a curve very close to $e_T$, only slightly more bound.
This suggests that there is no significant coupling between SO and
tidal contributions (as assumed in Eq.~\eqref{eq:Epieces}), even after contact. 

Finally, note that for $\ell\lesssim\ell_c$ the spin term $e_{SO}+e_{SS}$ is
probably influenced by hydrodynamical effects, so its correct
interpretation may be nontrivial. 
We also mention that similar results and conclusions
are obtained by using the EOB curves instead of the BBH data.

\subsection{Merger remnant}
\label{sec:dyn2}

\begin{figure}[t]
  \centering
    \includegraphics[width=0.49\textwidth]{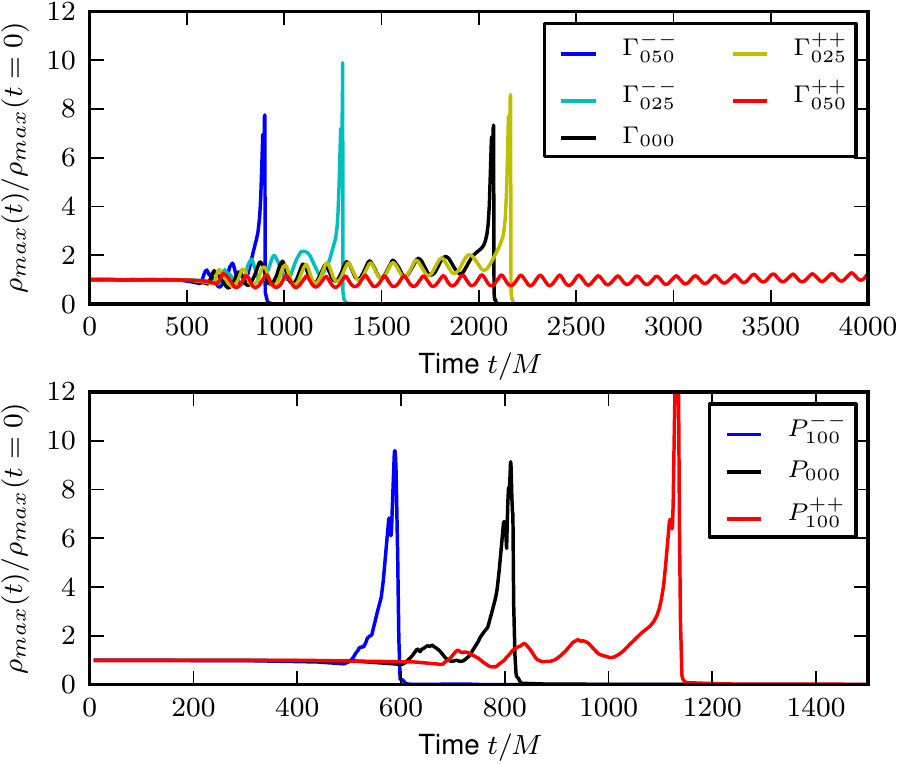}
    \caption{ \label{fig:rhomax} Evolution of the
      maximum mass-density $\rho(t)$ (normalized by its initial value)
      for the configurations
      $\Gamma_{050}^{++}$, $\Gamma_{025}^{++}$, $\Gamma_{000}$, 
      $\Gamma_{025}^{--}$ and $\Gamma_{050}^{--}$ (upper panel) and
      for the configurations $P_{100}^{++}$, $P_{000}$ and $P_{100}^{--}$ 
      (lower panel). Note the different $x$-axes.} 
\end{figure}

 \begin{figure}[t]
  \centering
     \includegraphics[width=0.49\textwidth]{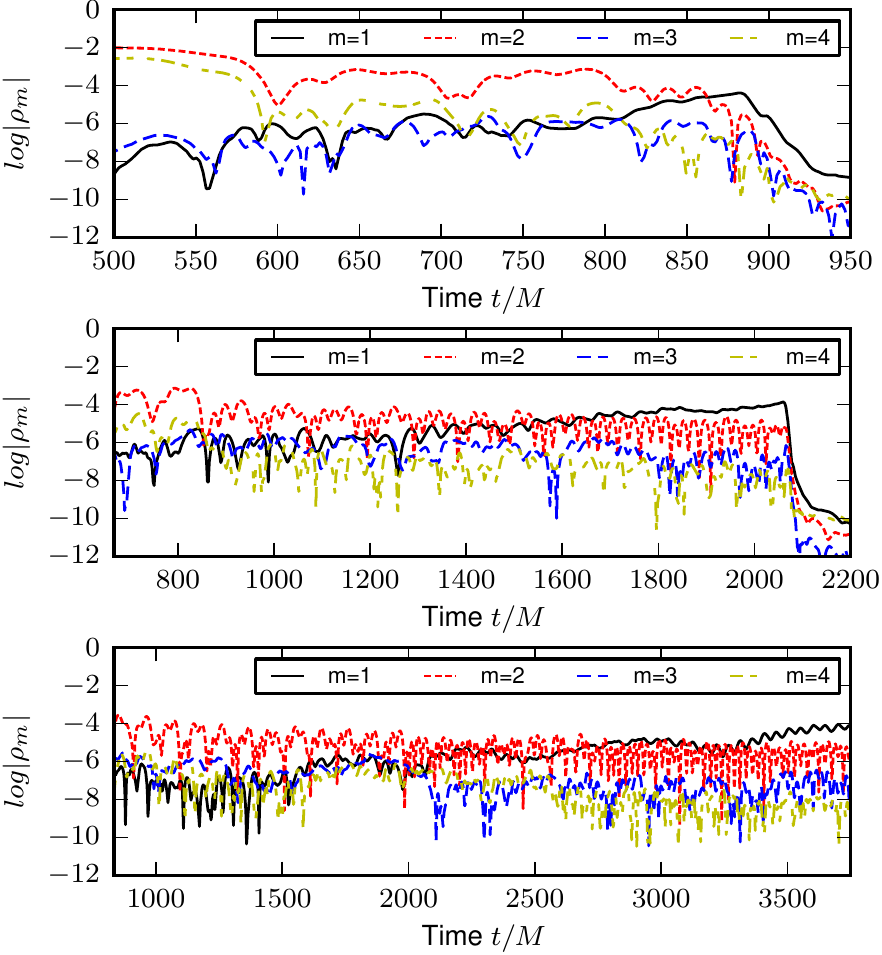}
    \caption{ \label{fig:rhomode} Evolution of projections $\rho_m(t)$
      for $m=1,2,3,4$ and different models. From top to bottom:
      $\Gamma_{050}^{--}$, $\Gamma_{000}$, $\Gamma_{050}^{++}$.}
\end{figure}

 \begin{figure*}[t]
  \centering
    \includegraphics[width=0.47\textwidth]{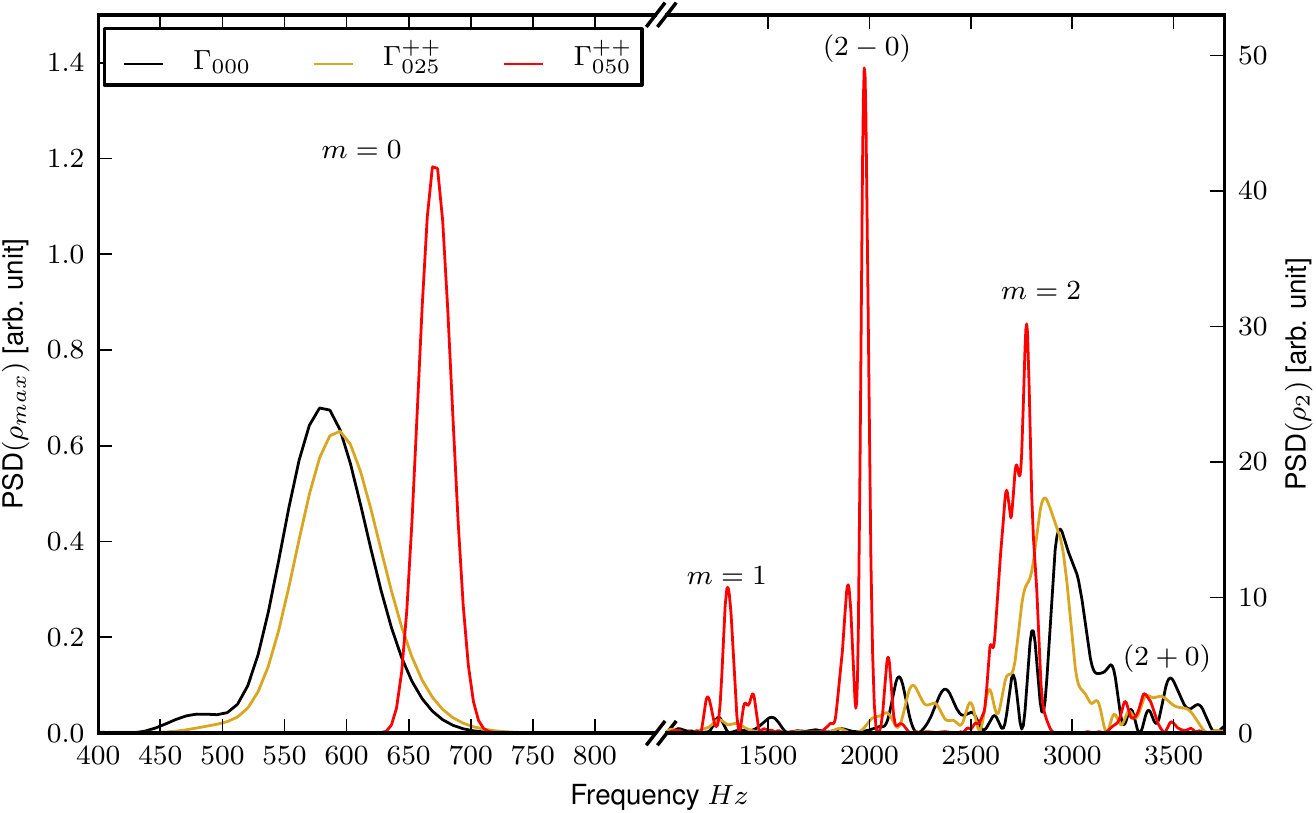} \quad
    \includegraphics[width=0.47\textwidth]{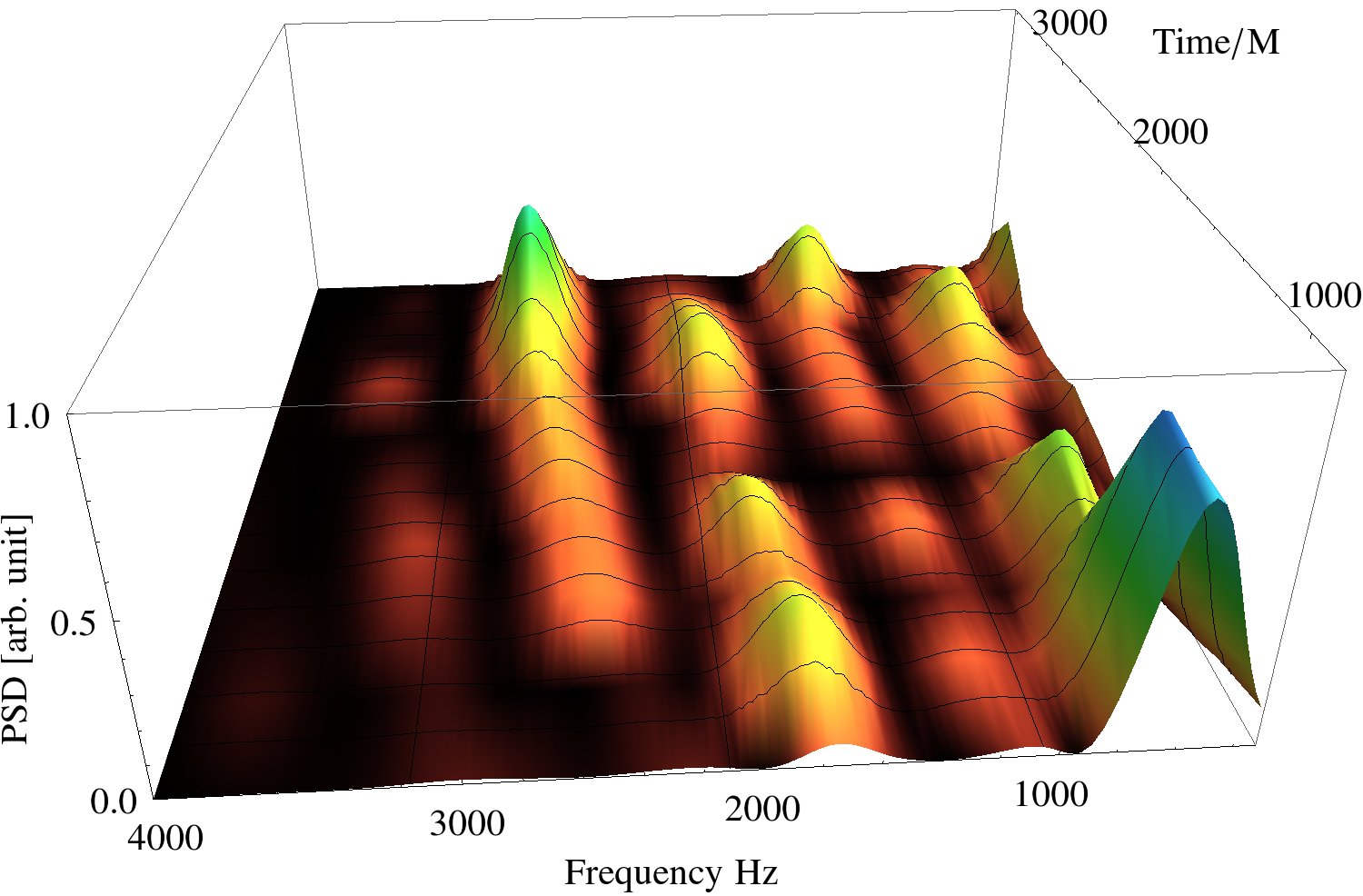}
    \caption{ \label{fig:fourier} 
      Fourier analysis of the rest mass projections $\rho_m$.
      Left: Power spectral density (PSD) of $\rho_{max}(t)$ and
      $\rho_2(t)$ for $\Gamma_{000}$, 
      $\Gamma_{025}^{++}$, $\Gamma_{050}^{++}$. 
      Right: Spectrogram of the quantity
      $\rho_{12}(t)\equiv\rho_1(t)+\rho_2(t)$ 
      in model $\Gamma_{050}^{++}$.}
\end{figure*}

All the configurations evolved with the $\Gamma$-law EOS
form, at merger, HMNSs characterized by different rotational states.  
In our simulations the HMNS is only supported by centrifugal forces
and thermal pressure (we include neither magnetic fields nor cooling
mechanisms). The angular momentum support is radiated away in GWs on
dynamical timescales, and the HMNS finally collapses. This happens
after about $1500$~M ($\sim22$~ms) from formation in the irrotational
model. The dimensionless angular momentum magnitude per reduced mass
of the HMNS is approximately $j\approx\ell_m\pm2\chi/\nu$ (assuming
$\chi\sim const$), e.g.~$j\sim3.18$, $3.53$, $3.9$ for $\Gamma_{050}^{--}$,
$\Gamma_{000}$, and $\Gamma_{050}^{++}$, respectively. 
We thus expect that configurations with
antialigned spins will collapse earlier, whereas configurations with
aligned spins will collapse later. 

In Fig.~\ref{fig:rhomax}
we show the evolution of the maximum rest mass density,
$\rho_{max}(t)$, for evolutions with the $\Gamma$-law EOS (upper panel) 
and the polytropic EOS (lower panel). The oscillations visible in the plot 
correspond to quasiradial modes (see below and Sec.~\ref{sec:gws}).
The average rest mass density increases linearly in time 
to about a critical density, $\rho_c\sim 2 \rho_{max}(t=0)$
($\rho_c\sim 1.2 \cdot 10^{15}$~$\text{g/cm}^3$), at which collapse
happens. As expected, we observe that model $\Gamma_{050}^{--}$ collapses
after approximately two quasiradial oscillations, model
$\Gamma_{025}^{--}$ after five, and  model $\Gamma_{025}^{++}$ after
twelve. Model $\Gamma_{050}^{++}$ survives for several dynamical timescales
and does not collapse until the end of the simulation ($t\sim 4000$~M.)
We have not evolved $\Gamma_{050}^{++}$ further since (i)~long-term
simulations can become inaccurate (see Appendix~\ref{app:tests}), and
(ii)~on these timescales other physical effects like magnetic fields
and neutrino cooling, presently not included, play an important role,
e.g.~\cite{Sekiguchi:2011mc,Paschalidis:2012ff,Hotokezaka:2013PhRvD..88d4026H,Deaton:2013sla}.  
However, considering a linear trend in $\rho_{max}(t)$, we
extrapolate that collapse should happen at about $\sim167000$~M
($\sim272$~ms) after merger. 

The lower panel of Fig.~\ref{fig:rhomax} refers to configurations
evolved with a polytropic EOS. 
Since in this case thermal pressure support is absent, collapse 
occurs much earlier than for the $\Gamma$-law EOS.
The HMNS of the irrotational configuration collapses after about 
one quasiradial oscillation; model $P_{100}^{--}$ promptly collapses without
HMNS formation, and model $P_{100}^{++}$ collapses after few
oscillations. 

During its evolution the HMNS oscillates nonlinearly and becomes
progressively more compact. The oscillations modes can be identified
as the quasiradial mode, the $m=2$ $f$-mode, and nonlinear
combinations of them, e.g.~\cite{Shibata:1999wm,Stergioulas:2011gd}.
A way to characterize nonlinear modes is to project the rest-mass
density onto spherical harmonics, e.g.~\cite{Baiotti:2008nf}. For
simplicity, we consider $\rho(x,y,z=0,t)$ in the orbital plane $z=0$ and
the projections~\cite{Stergioulas:2011gd} 
\be
\label{eq:rhom}
\rho_m(t) = \int  \rho(x,y,z=0,t) e^{i m \phi(x,y)} \text{d}x \text{d}y \ .
\ee
In Fig.~\ref{fig:rhomode} we report the evolution of the first
projections $m=1,2,3,4$ for some representative runs. 
The figure shows that the dominant mode is the $m=2$ mode. 
Actually, the projection/mode with larger amplitude is the
quasiradial one ($m=0$, also visible in $\rho_{max}$
Fig.~\ref{fig:rhomax}). As we shall discuss later, however, this mode
has a frequency too low to be visible in the GW spectrum.
The evolution of $\rho_m(t)$ is qualitatively similar in the 
different configurations, with differences only related to the
collapse time. A strong $m=1$ mode appears in \emph{all} simulations 
before collapse (see e.g.~the central panel) and also
dominates the evolution of $\Gamma_{050}^{++}$ after $t\sim3500$~M. 
We interpret it as a physical hydrodynamical effect due to mode
couplings, but we cannot rule out that it is triggered by some
numerical effect.

\begin{table}[t]
  \centering  
  \caption{ \label{tab:fpeak}     
    Peak frequencies of the power spectral density (PSD) of $\rho_m$
    and $\rho_{max}$. They are estimated by fitting a Gaussian of
    standard deviation $\sigma$. The value of the latter 
    is reported in parenthesis.}
  \begin{tabular}{cccc}        
    \hline
    & $m=0$ & $m=1$ & $m=2$ \\
    \hline  
    $\Gamma_{000}$      & 584 (34) & 1543 (38) & 2974 (114) \\
    $\Gamma_{025}^{++}$ & 594 (34) & 1482 (38) & 2871 (103) \\      
    $\Gamma_{050}^{++}$ & 671 (13) & 1341 (13) & 2738 (76) \\
   \hline
  \end{tabular}
\end{table}

In order to extract the mode frequencies, we perform a Fourier
analysis of the $\rho_m$ projections and $\rho_{max}$. The
quasiradial mode is best extracted from the latter. The Fourier
transform is performed only in the part of the signal after merger,
i.e.~$t>t_m$. 
Some of the relevant results are summarized in
Fig.~\ref{fig:fourier}, where we show on the left the spectra of
$\rho_{max}$ for lower frequencies and $\rho_{2}$ for higher frequencies for 
different models and on the right the spectrogram of
model $\Gamma_{050}^{++}$. Focusing on the left panel, we observe that
the spectrum is composed of few frequencies;  we identify $m=0,1,2$ modes
together with nonlinear couplings
``$2\pm0$''~\cite{Stergioulas:2011gd,Dimmelmeier:2005zk,Baiotti:2008nf}. For
model $\Gamma_{050}^{++}$ the highest power is actually found at the
``2-0'' frequency in $\rho_2$.

The peak frequencies for the different modes and models are stated in
Tab.~\ref{tab:fpeak} for the relevant case of spin aligned with the
orbital angular momentum.  The frequency peak of the $m=2$ mode
becomes larger the smaller the HMNS rotation is. This is because the
HMNS with more angular momentum support is less compact, and the proper
frequencies decrease if the compactness decreases (compare with
sequences of a single rotating star with the same rest mass
in~\cite{Dimmelmeier:2005zk}). 
Notably, for model $\Gamma_{050}^{++}$ the observed frequency shift
with respect to the irrotational configuration is 236~Hz. The value is
significant at the $1$-$\sigma$ level, see Tab.~\ref{tab:fpeak}. 
Differently from the $m=2$ mode, the frequency of the quasiradial
mode ($m=0$) is found to increase for HMNS with larger angular
momentum. As discussed in~\cite{Stergioulas:2011gd}, the quasiradial
mode frequency depends on the compactness of the HMNS and on how close
the star model is to the collapse-instability threshold.
The larger the compactness, the larger the mode frequency is; but
configurations close to the instability threshold can have smaller
frequencies since the instability threshold is a neutral point. 
We interpret our results according to the above argument: HMNSs with
larger angular momentum support are further from collapse threshold and thus
have higher frequencies. 

The spectra lines appear broad due to the highly dynamical nature of 
the HMNS. Investigating the dynamical excitation of the modes by a
spectrogram, we find that 
(i)~the modes are ``instantaneously'' characterized by relatively narrow peaks; 
(ii)~different modes dominate different parts of the signal;
(iii)~some of the peaks ``drift'' towards higher frequencies as the
HMNS becomes more compact.
The right panel of Fig.~\ref{fig:fourier} shows the spectrogram of
the quantity $\rho_{12}(t)\equiv \rho_1(t)+\rho_2(t)$
for $\Gamma_{050}^{++}$.
At early times the $m=0$ (quasiradial) mode dominates the
$\rho_{12}$ spectrum, but around $t\sim2000$~M the $m=2$ becomes
the main oscillation 
mode.  A ``drift'' of the $m=2$ mode towards higher frequencies is visible,
which corresponds to the fact that the HMNS becomes more compact. 
The ``2-0'' coupling remains the secondary peak during the whole
simulation. The $m=3$ and ``2+0'' modes are also visible. At the very end of
the evolution the $m=1$ mode has the largest power.

\begin{table}[t]
  \centering  
  \caption{ \label{tab:remnant}     
      Important quantities for the merger remnant. Stated are the black hole mass, the dimensionless spin of 
      the black hole, and the absolute disk mass of the surrounding disk as well as the percentage with respect 
      to the total baryonic mass. }
  \begin{tabular}{l|cccc|ccc}        
    \hline
    &  $\Gamma_{050}^{--}$ & $\Gamma_{025}^{--}$ & $\Gamma_{000}$ & $\Gamma_{025}^{++}$ & P$_{100}^{--}$ & P$_{000}$ & P$_{100}^{++}$ \\
    \hline  
    $M_{BH}$             & 2.92  & 2.88  & 2.85  &  2.86  & 2.95  & 2.94  & 2.89 \\ 
    $\chi_{BH}$          & 0.80  & 0.79  & 0.78  &  0.79  & 0.81  & 0.83  & 0.84 \\
    $M_{\rm b,\ disk}$   & 0.039 & 0.068 & 0.081 &  0.082 & 0.006 & 0.021 & 0.065 \\
    $M_{\rm b,\ disk}/M_b$ & 1.2\% & 2.1\% & 2.5\% &  2.5\% & 0.2\% & 0.6\% & 2.0\% \\   
   \hline
  \end{tabular}
\end{table}

Finally, we briefly discuss the black hole and the remnant disk.
All simulations (except $\Gamma_{050}^{++}$) result in a
black hole surrounded by a nonmassive accretion disk. 
Tab.~\ref{tab:remnant} summarizes the irreducible mass and the 
dimensionless spin of the black
hole, and the rest mass of the disk. 
The black hole mass is larger for antialigned spin configurations and spin
magnitude, and a monotonic trend is observed for smaller spin and aligned
spin configurations. The opposite holds for the disk mass. 
The spin of the black hole is larger for 
aligned configurations in barotropic evolutions. 
This effect is not visible in the $\Gamma$-law simulations, in which
the more massive disk probably has also larger angular momentum. 
The maximum spin produced is $0.84$, which is consistent
with the upper limit found in~\cite{Kastaun:2013mv}.
Notice that all reported quantities are affected by large uncertainties, and
they should be considered only as a qualitative indication.
For example the uncertainty on the black hole mass calculated from
$L_2$ and $H$ runs of  the irrotational configuration is $\sim 0.01$.

\section{Gravitational radiation}
\label{sec:gws}

\begin{figure}[t]
  \centering
    \includegraphics[width=0.49\textwidth]{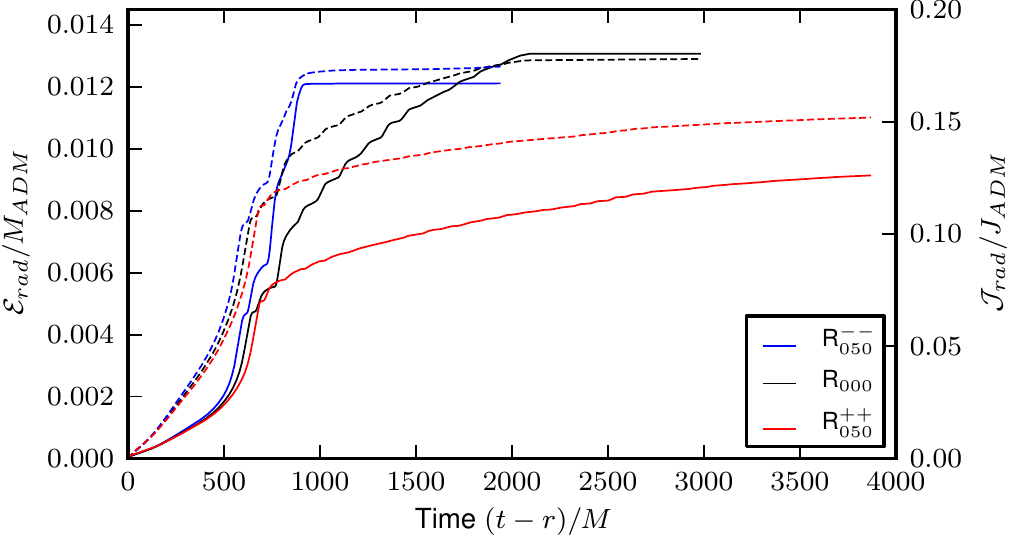}
    \caption{ \label{fig:EJRad} Energy (solid lines) and angular
      momentum (dashed lines) radiated in GWs for models
      $\Gamma_{050}^{--}$,$\Gamma_{000}$, and $\Gamma_{050}^{++}$.}
\end{figure}

\begin{figure*}[t]
  \centering
    \includegraphics[width=0.48\textwidth]{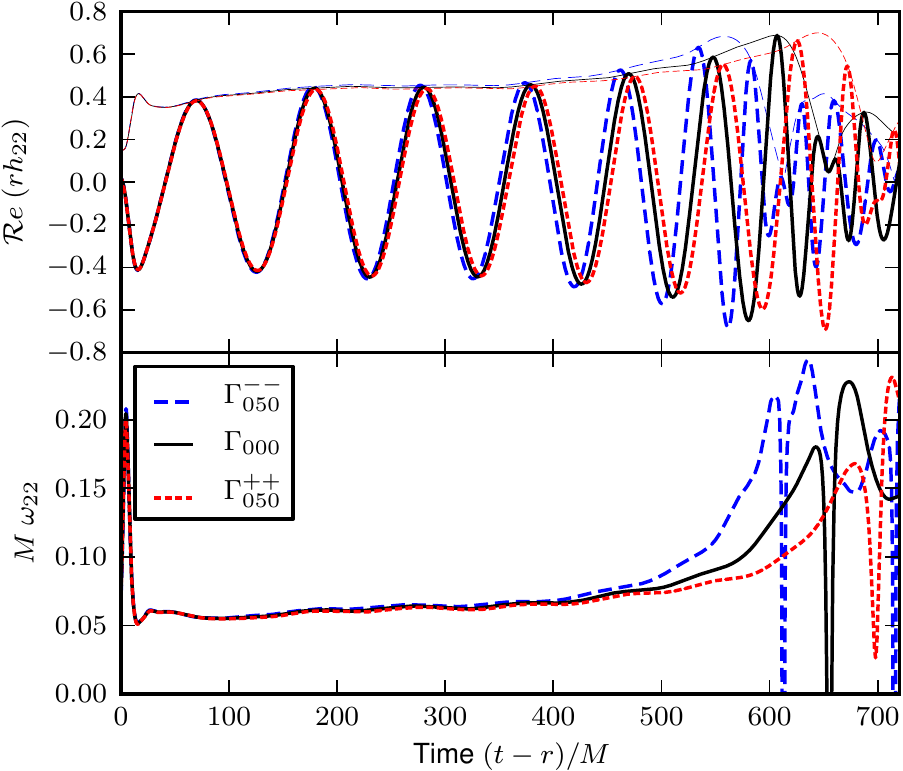} \quad
    \includegraphics[width=0.48\textwidth]{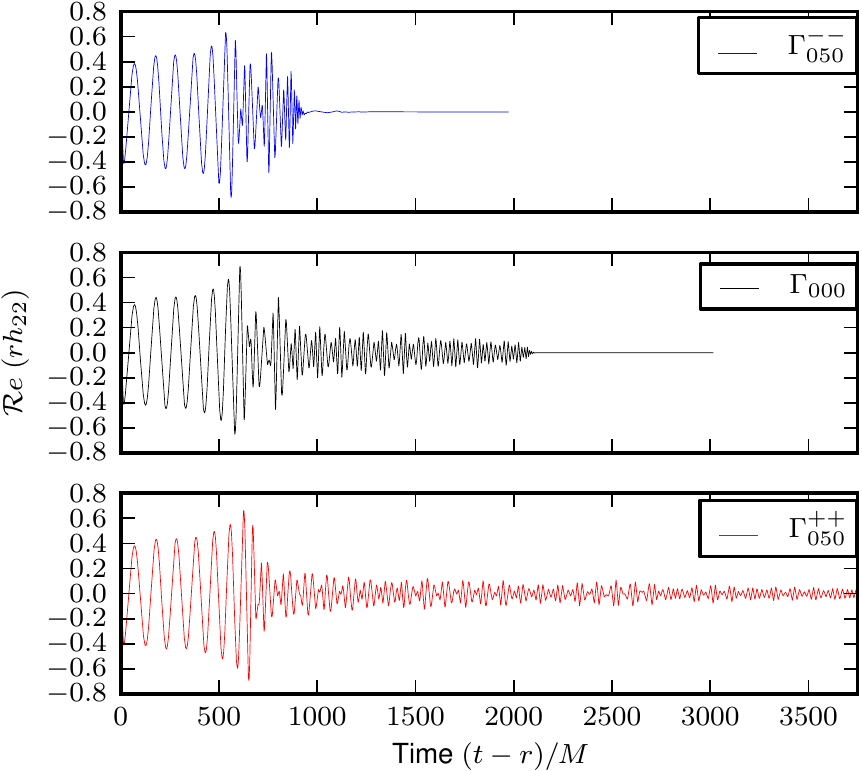} 
    \caption{ \label{fig:waves} Gravitational wave
      signal for models $\Gamma_{050}^{--}$,
      $\Gamma_{000}$, and $\Gamma_{050}^{++}$. 
      Left: Inspiral waveforms $\Re{(r\, h_{22})}$ and $r|h_{22}|$,
      and frequency $M \omega_{22}$.
      Right: Full signal $\Re{(r\, h_{22})}$.} 
\end{figure*}

\begin{figure}[t]
  \centering
    \includegraphics[width=0.49\textwidth]{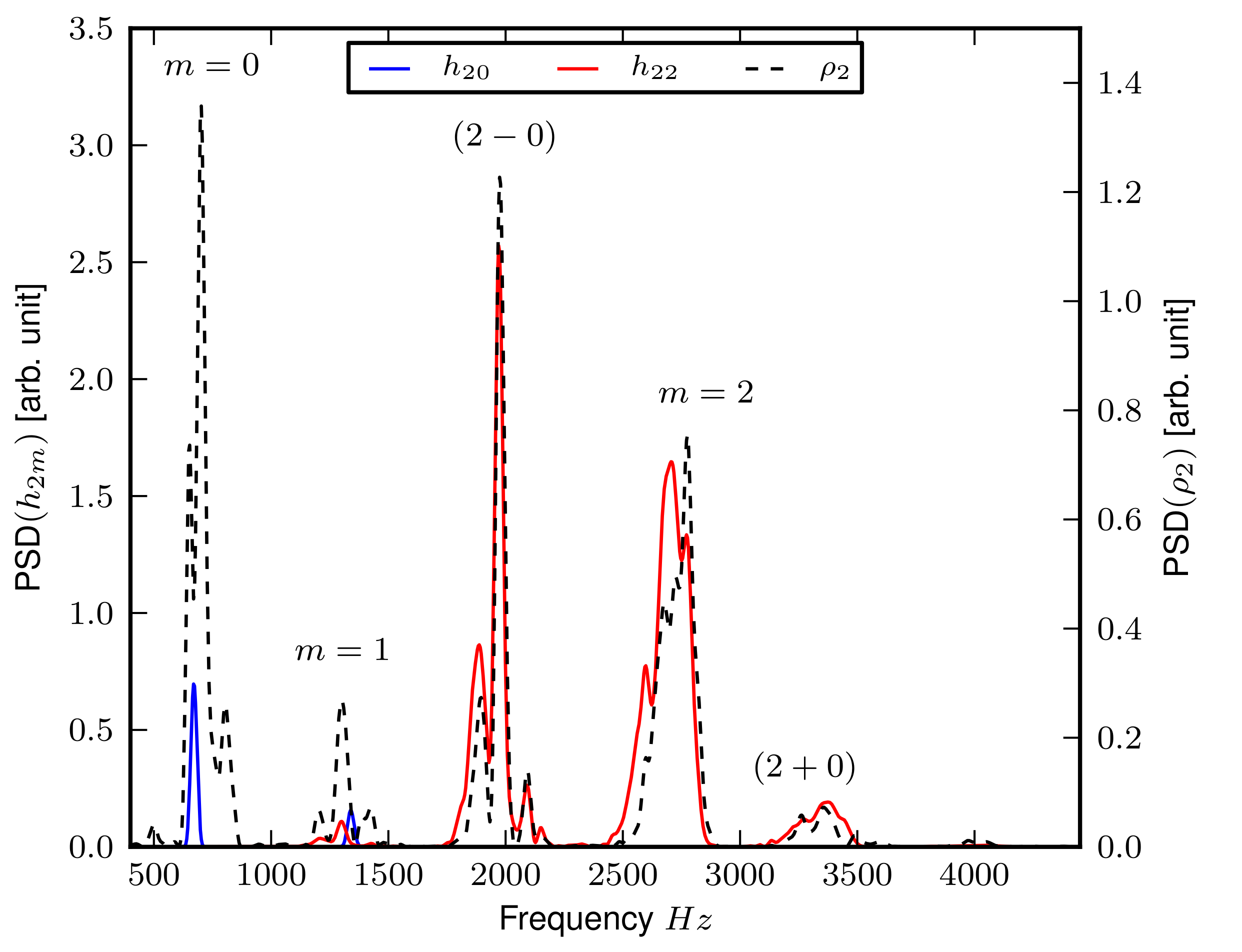}
    \caption{ \label{fig:HMNSgw} Fourier analysis of the $l=2$
      postmerger waveform multipoles and matter projection $\rho_2$
      for model $\Gamma_{050}^{++}$. The waveform frequencies strongly
      correlate with the fluid's modes.} 
\end{figure}

The dynamics described in Sec.~\ref{sec:dyn} is relatively simple (but
far from trivial). For sufficiently high spin magnitudes,
$\chi\sim0.05$, the SO interaction is a significant repulsive
(attractive) contribution for aligned (antialigned) spins
configurations. For aligned configuration the SO competes with finite
size effects. At merger, however, binaries with aligned spins are more
bound. HMNSs are formed with more or less angular momentum support
than in the irrotational configuration ($j\sim\ell_m \pm 2\chi/\nu$);
thus they are either closer or farther from the collapse threshold
(radial instability point). We discuss in this section how the emitted
gravitational radiation encodes all this.

The total energy and angular momentum emitted in GWs quite differ in
the different models, as can be seen from Fig.~\ref{fig:EJRad}. 
The irrotational configuration emits about $1.2\%$ of the initial ADM mass
and $18\%$ of the initial angular momentum. $\Gamma_{050}^{--}$ emits about
the same amount, but in about half the time. To the end of the
simulation, $\Gamma_{050}^{++}$ has emitted $0.8\%$ of the initial mass and
about $15\%$ of the initial angular momentum.  In all the cases, the
main emission channel is the $l=m=2$ multipole that alone accounts for
$\sim97\%$ of the emitted energy. However, in the postmerger phase
other channels are clearly excited; the largest amplitudes are
observed in the $l=2$, $m=0$, the $l=3$, $m=3,2$, and the $l=m=4$ modes 
(in that order).

Figure~\ref{fig:waves} (left panel) shows the $l=m=2$ inspiral waveforms,
focusing again on the models $\Gamma_{050}^{--}$, $\Gamma_{000}$ and $\Gamma_{050}^{++}$
for clarity. 
Intermediate results are of course found for the other models.  
The upper-left panel shows the real part and amplitude of the $l=m=2$ mode
of the GWs, the lower-left panel the GW frequency
$M\omega_{22}=-\Im(\dot{h}_{22}/h_{22})$; note the retarded time in the $x$-axis.
The merger times, computed at the peak of $|rh_{22}|$, are
$t_m\sim595,\ 551,\ 636 M $ for $\Gamma_{000}$, $\Gamma_{050}^{--}$ and
$\Gamma_{050}^{++}$, respectively (see also
Tab.~\ref{tab:dynamics}). The peaks of the wave amplitude are all very
close to $\sim0.7$. 
The GW frequency corresponding to the peak of the wave amplitude is
smallest for aligned spin. 
At merger $M\omega_{22}\sim0.127,0.124,0.123$ 
for $\Gamma_{000}$, $\Gamma_{050}^{--}$, and $\Gamma_{050}^{++}$, respectively. 
At contact instead $M\omega_{22}\sim0.069,0.067,0.071$. Note that
these frequencies have uncertainties of about $10\%$. 
The effect of spin-orbit  interaction is clearly visible from the plot. 
Computing the accumulated phase of the GW, we find that the
irrotational configuration emits 7.0 GW cycles to merger;
$\Gamma_{050}^{--}$ emits 6.3 cycles and $\Gamma_{050}^{++}$ emits 7.3 cycles. 
This phase difference results from the dynamics discussed in
Sec.~\ref{sec:dyn1}, and encodes the interplay of spins and tidal
interactions. 

Let us finally discuss the emission from the
HMNS. Figure~\ref{fig:waves} (right panel) shows the $l=m=2$ complete
waveform. The earlier the HMNS collapses, the larger the amplitude
of the wave in the postmerger phase is. 
As also shown in
Fig.~\ref{fig:EJRad}, model $\Gamma_{050}^{--}$ emits more energy
and angular momentum than $\Gamma_{000}$ and $\Gamma_{050}^{++}$ during the
first $\sim600$~M after merger.  
In order to identify the origin of the emission, we perform a Fourier
analysis of the $l=m=2$ and $l=2$ $m=0$ multipoles, and compare this with
the mode analysis of Sec.~\ref{sec:dyn2}. As in the previous section,
we consider only the signal at $t>t_m$. 
A relevant example of this analysis is summarized in
Fig.~\ref{fig:HMNSgw}. The spectra of the waves and matter
modes strongly correlate: the HMNS modes are the main emitters during
the postmerger phase~\cite{Stergioulas:2011gd}. 
We stress that the complete GW spectrum includes also the inspiral part
of the GW signal. In particular, the merger happens at GW frequencies of $\sim
1.2-1.3$~kHz, and, up to these frequencies, the spectrum is dominated
by the inspiral. Thus the quasiradial mode frequency of the HMNS is
not observable, whereas the $m=2$ and ``$2\pm0$'' peaks form the main
postmerger signal.   

In~\cite{Oechslin:2007gn,Bauswein:2011tp} (see
also~\cite{Hotokezaka:2013PhRvD..88d4026H}) it is shown that the frequency of the
peak of the GW (postmerger) spectrum is strongly dependent on the
EOS, and, to a lesser extent, on the 
total mass, mass ratio and spin. The latter aspect has been investigated by
comparing irrotational and corotational configurations for a few
models, and no significant frequency shift was observed.
The long wave train of model
$\Gamma_{050}^{++}$ allows us to resolve a significant frequency shift,
suggesting that spin effects may be more important 
than previously thought. Note that a shift towards \emph{lower}
frequencies can favor GW detection by advanced interferometers.

\section{Conclusion}
\label{sec:conc}

We have studied BNS mergers in numerical relativity with a
realistic prescription for the spin. Consistent initial data have been
produced with the CRV approach and evolved for the first time. 

We have considered moderate star rotations corresponding to
dimensionless spin magnitudes of $\chi=0.025, 0.05$, 
and direction aligned or antialigned with the orbital angular
momentum. The dimensionless spins $\chi$ are estimated by considering
the angular momentum and masses of stars in isolation with the same
rotational state as in the binary. We have investigated the orbital
dynamics of the system by means of gauge invariant $\E(\ell)$
curves~\cite{Damour:2011fu}.  

Our simple proposal for the estimation of $\chi$ proved to be robust,
and allows us to show consistency with PN and EOB energy curves at
early-times. Using energy curves we have also compared, for the 
first time to our knowledge, BNS and BBH dynamics
(see~\cite{Foucart:2013psa} for a waveform-based comparison of the
case BBH--mixed binary.) We extracted and isolated different
contributions to the binding energy, namely the point-mass nonspinning
leading term, the spin-orbit and spin-spin terms and the tidal term.
The analysis indicates that the spin-orbit contribution to the binding
energy dominates over tidal contributions up to contact (GW frequencies
$M\omega_{22}\sim0.07$) for $\chi\sim0.05$. The
spin-spin term, on the other hand, is so small that it is not well resolved in
the simulations. No significant couplings between tidal and spin-orbit
terms are found, even at a stage in which the simulation is in the
hydrodynamical regime (at this point, however, the interpretation of
``spin-orbit'' probably breaks down.)

The spin-orbit interactions significantly change the GW signal
emitted. During the three orbit evolution we observe an accumulated
phase differences up to $0.7$ GW cycles (over three orbits) between
the irrotation configuration and the spinning ones ($\chi=0.05$), that
is we obtain first quantitative results for orbital ``hang-up'' and
``speed-up'' effects. A precise modeling of the late-inspiral-merger
waveforms, as in~\cite{Bernuzzi:2012ci}, needs to include spin effects
even for moderate magnitudes.   
Long-term (several orbits)
simulations are planned for a thorough investigation of this aspect,
together with detailed waveform phasing analysis and comparison with
analytical models. Extensive simulations with different EOSs will be
also important to check the universal relations recently proposed
in~\cite{Bernuzzi:2014kca}. 

We have also investigated spin effects on the formation and collapse of the
merger remnants (HMNSs), and the hydrodynamical evolution the HMNS
modes~\cite{Stergioulas:2011gd}. The star rotation influences the   
HMNS produced at merger by augmenting (aligned spin configuration) or
reducing (antialigned) the angular momentum support. Earlier or
delayed collapse of several milliseconds is thus observed depending
on the spins orientation.
We have found that characteristic frequencies of
the HMNS are shifted to \emph{lower} values by rotation. This suggest
that spin effects may be more important 
than previously thought. HMNS modes are the main emitters of GWs in the
postmerger phase, and they may allow for a precise determination of the
neutron star radius in a GW detection~\cite{Bauswein:2011tp}. Extensive
evolutions of CRV configurations for various EOSs and spins are 
needed in order to assess the role of spin and to obtain accurate 
phenomenological relations for frequency vs.\ radius.

Future work should also be devoted to understanding the impact of our result on 
GW astronomy. 
We expect that some aspects of spin in BNS can be modeled similarly to the GW
analysis for nonprecessing spinning  
BBHs~\cite{Ajith:2009bn,Hannam:2010ec,Santamaria:2010yb,Hannam:2007wf}.
Furthermore, it would be important to explore the
relevance of spin-orbit corrections in the construction of templates
for detecting the star's EOS~\cite{Damour:2012yf,Favata:2013rwa},
possibly applying realistic data analysis
settings~\cite{DelPozzo:2013ala}. 
In the relevant case of aligned spin configurations, spin-orbit
effects actually compete with finite size effects.
One might expect that, for some realistic spin magnitudes, this could
affect the measurability of the 
EOS (tidal polarizability parameters) when spin is not properly taken into
account. Similarly, if the spin is estimated from the early
inspiral, a bias in the spin magnitude could 
significantly affect the measure of the tidal
parameters~\cite{Damour:2012yf}.

\begin{acknowledgments}
  It is a pleasure to thank Alessandro Nagar for valuable comments and
  interactions, and Simone Balmelli for providing us with the EOB
  curves. We are also grateful to Nathan Kieran Johnson-McDaniel and 
  Andreas Weyhausen for interesting discussions.
  This work was supported in part by DFG Grant SFB/Transregio~7
  ``Gravitational Wave Astronomy'', the Graduierten-Akademie Jena,
  and NSF Grants PHY-1204334 and PHY-1305387.
  Simulations were performed on SuperMUC (LRZ) and JUROPA (JSC). 
\end{acknowledgments}

\appendix

\section{Single spinning stars}
\label{app:singlestar}

In the CRV approach one assumes the existence of an approximate
helical Killing vector. In an inertial frame it has the 
form~\cite{Tichy:2012rp}
\begin{equation}
\label{xi_inertial}
\xi^{\mu} = 
\left( 1, -\Omega [x^2 - x^2_{CM}], \Omega [x^1 - x^1_{CM}], 0  \right) .
\end{equation}
Here $x_{CM}^i$ denotes the center of mass position of the system,
and $\Omega$ is the orbital angular velocity, which we have chosen to lie
along the $x^3$-direction.

For a single star $x_{CM}^i$ coincides with the star center $x_{C*}^i$.
Furthermore if we follow the CRV approach $\Omega=0$, since a single star is
not orbiting. Thus the approximate Killing vector simply points along the
time direction.
We can then set the $\omega^j$ in Eq.~\eqref{w_choice} to the same value as in
the case of binary stars. If we now solve the CRV equations we obtain
initial data for a single spinning star. This spin can be unambiguously
computed from the ADM angular momentum and reported in the $S_s$ column
of Tab.~\ref{tab:models}.

However, there is at least one other way to obtain single spinning stars.
We can set $\Omega$ to a nonzero value and assume that the approximate
Killing vector is truly helical. If we then assume that the fluid velocity
is along the Killing vector
\be
{u}^{\mu} = u^0 \xi^{\mu} ,
\ee
we obtain the standard assumptions for a corotating configuration, but for a
single star only. If we now solve the usual equations for the corotating
case (see e.g.~\cite{Tichy:2009yr}), we also obtain
initial data for a single spinning star. Notice however, that the star spin
in this corotating approach is about 10\% higher than in the CRV approach
if we set $\Omega=\omega$. This means that $\Omega$ and $\omega^j$ do not
have the same meaning, which is not too surprising considering that 
$\omega^j$ is just an auxiliary local field in the CRV construction,
while $\Omega$ is the angular velocity seen by observers at infinity.

The above observations can be used to estimate the angular velocity seen by
observers at infinity when using the CRV approach for single stars. We first
construct a single star using the CRV approach for a particular $\omega^j$
and compute its spin $S_s$. We then choose $\Omega$ such that the corotating
approach results in the same spin. We can then interpret this $\Omega$ as
the angular velocity seen by observers at infinity for a single star with
spin $S_s$. If we follow these steps for e.g. $\omega^z=0.0046$, we find
that we have to choose $\Omega=0.0042$ to obtain the same spin with the
corotating approach. Thus the angular velocity seen at infinity for
$\omega^z=0.0046$ is really only $0.0042$, which makes sense considering
that any local frequency will be redshifted by the time it is observed at
infinity. Thus the spin period observed at infinity is about 10\% larger
than what we get from $2\pi/\omega^z$.

\section{Equilibrium sequences}
\label{app:IDseq}

\begin{figure}[t]
  \centering
    \includegraphics[width=0.5\textwidth]{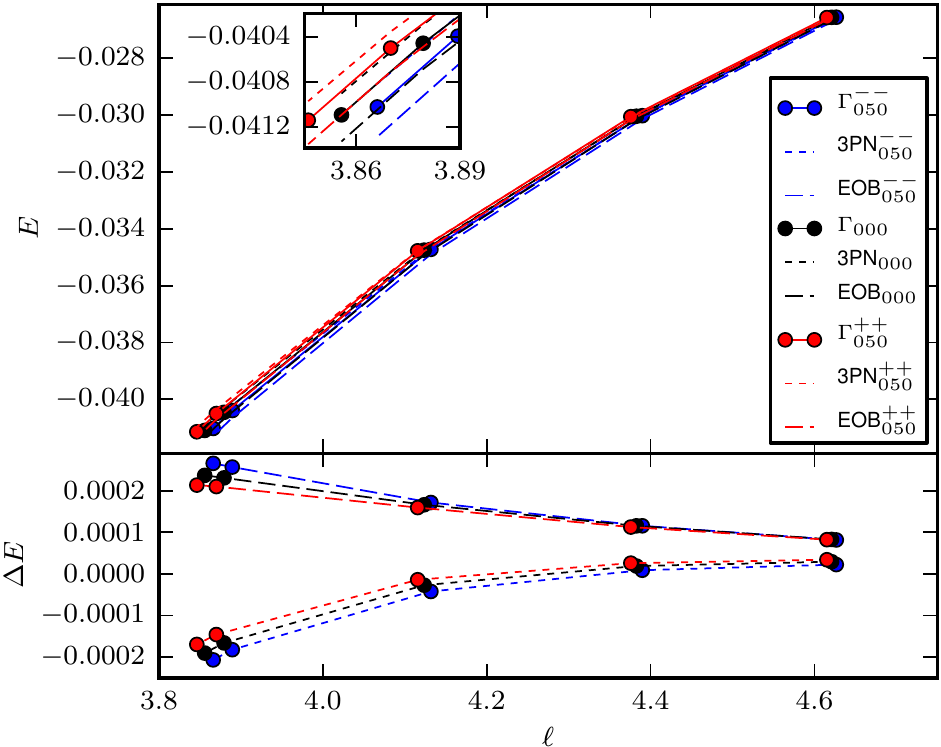}
    \caption{ \label{fig:Ej_equil} 
      Top: Binding energy vs.\ orbital angular
      momentum curves for equilibrium
      configurations together with 3PN and EOB results.
      Bottom: The  
      differences $\Delta \E = \E - \E^{X}$ with $X=$3PN, EOB.}
\end{figure}

In this appendix we present the equilibrium sequences of CRV data,
considering in particular the curves $E(\ell)$, where $\ell=(
J_{\rm ADM} - 2 S_s )/(M_s^2\nu)$ and $E = (M_{\rm 
  ADM}/M_s-1)/\nu$; see also Eq.~\eqref{Ebdef} and~\eqref{jdef}. 
The numerical data are again compared to PN and
EOB~\cite{Buonanno:1998gg} results, as described in the main text.

In Fig.~\ref{fig:Ej_equil} we report the curves $E(\ell)$ for
sequences with $\omega^z=0,\pm0.0023$. 
The 3PN and adiabatic EOB curve are very close to the data for large
separations. The differences are quantified in the right panel, by
plotting $\Delta E = E - E^{X}$ with $X=$ 3PN, EOB. 
For the closest separation computed, the sequences equally deviate
from EOB and 3PN curves, but while the 3PN result predicts a less bound
binary, the EOB method predicts a more bound one. Note also a systematic
difference in $\Delta\E$ for different spins.

\section{Robustness of simulations in the postmerger phase}
\label{app:tests}

The accuracy of the simulations in the orbital phase has been studied
in different recent works. In particular
Refs.~\cite{Thierfelder:2011yi,Bernuzzi:2011aq} presented the 
first convergence tests of waveform's phase and amplitude in three and
nine-to-ten orbits inspirals. We do not repeat that analysis here. The
same works pointed out that after merger convergence cannot be
monitored in the waveforms, and, in general, the results are much more
dependent on the resolution and grid setup employed.
See also~\cite{Baiotti:2009gk} for similar conclusions obtained with
other codes.  
In this appendix we discuss the robustness of the simulations in the
postmerger phase, in particular regarding the merger remnant,
i.e.~HMNS. We consider two 
different series of tests: (i)~an {\it internal} test based on a resolution
study and different grid setup, and (ii)~an {\it external} test that
compares the same evolution of similar initial data obtained with
SGRID and Lorene. We focus on the irrotational configuration.

\begin{figure}[t]
  \centering
  \includegraphics[width=0.49\textwidth]{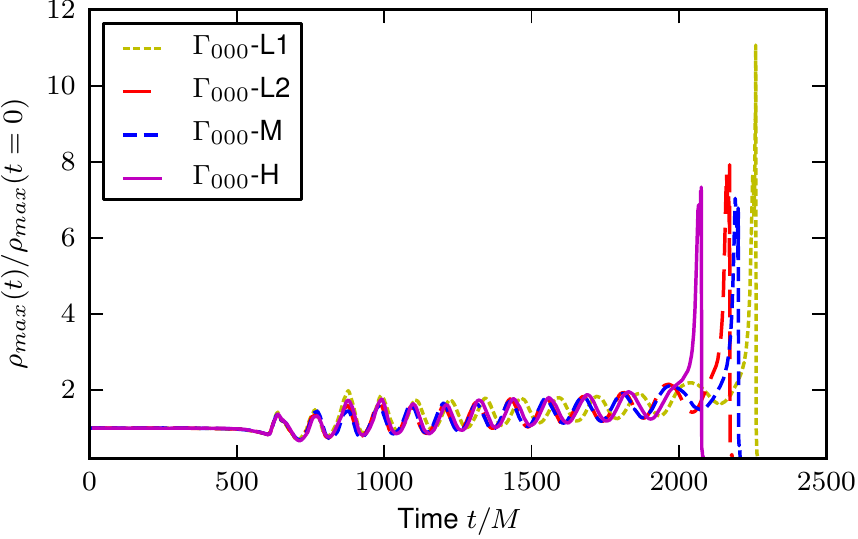}
  \caption{ \label{fig:rhomax_res} Evolution of the
    maximum rest mass density $\rho(t)$ (normalized by its initial value) for 
    $\Gamma_{000}$ using different resolutions.}
\end{figure}

\begin{figure}[t] 
  \centering
    \includegraphics[width=0.48\textwidth]{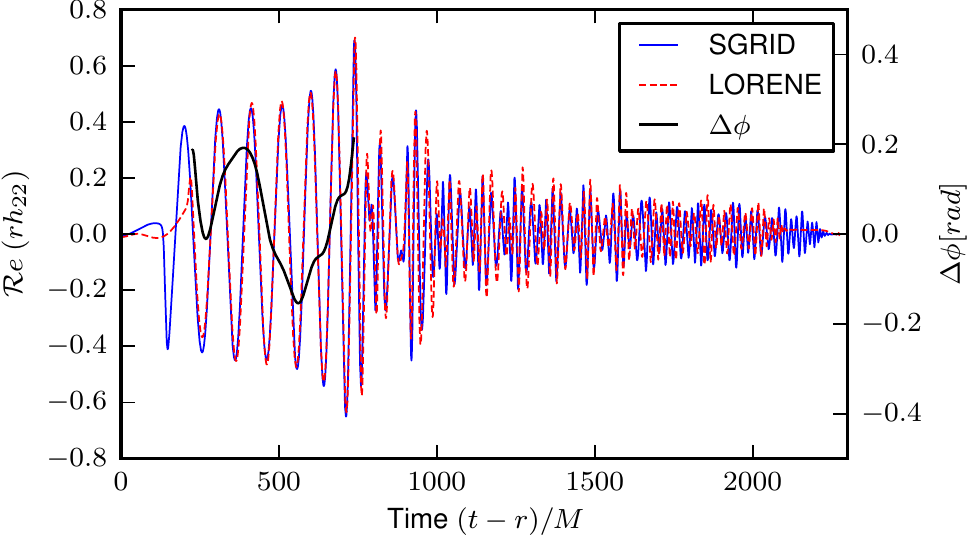} 
    \caption{\label{fig:loreneVsgrid} GWs from runs with SGRID
      and Lorene initial data. Note the two initial configurations have
      different separations. Shown is the $l=m=2$ mode, the Lorene
      data are suitably shifted for the comparison. The black line
      refers to the phase difference up to merger.}
\end{figure}

Fig.~\ref{fig:rhomax_res} shows the evolution of the
maximum rest-mass density on the finest refinement level for the
different resolutions considered in this work (see Tab.~\ref{tab:grid}). 
The results show a converging behavior of this quantity
with increasing resolution, making us confident that the chosen setup
gives, at least qualitatively, correct results. 
As observed in previous works, it is impossible to prove strict convergence
either in this quantity or in the waveforms.

Extensive tests in an early stage of the work have shown that the
nonconservative mesh-refinement of BAM is not optimal for long-term
evolution of the HMNS. During the inspiral the compact stars are
contained and completely resolved in a single Cartesian box at the
finest refinement level. In the postmerger phase, however, a
significant amount of matter \emph{can} cross grid boundaries,
unsurprisingly leading to severe violations of the rest mass
conservation.  Only when the inner box encloses most matter, we expect
systematic convergence.

As an example we consider the grid configuration L2 and an
equivalent configuration in which the number of points in the moving
levels are reduced from $n^{\rm mv}=144$ to $n^{\rm mv}=96$ (but the
same resolution is used). With smaller boxes the outer layer of the
HMNS are not covered by the finest refinement level.
The larger mass violation of the setup with $n^{\rm mv}=96$ led to
\emph{earlier} (in this case study) black hole formation by about
$\Delta t\sim700$~M.   
Note however that the rest mass is conserved for the L2 grid up to
$\sim2\%$ to collapse, while for the H grid up to $\sim0.8\%$. 

In a second series of tests, we compare the evolutions obtained with
the SGRID initial data with Lorene data~\cite{LORENE}. 
The Lorene data considered here have been
employed in several works in the past, e.g.~\cite{Baiotti:2009gk,Thierfelder:2011yi}.
The initial separation slightly differs in the two cases: the proper
distance is $\sim13.3$~M for SGRID data and $\sim13.0$~M for Lorene data.  
Lorene data employ four domains and the number of collocation points for
each domain is $N=33\times25\times24$.
SGRID uses 4 compactified domains with $N=24\times24\times8$ points 
and 2 Cartesian domains with $N=20\times20\times20$.
The grid configuration used for the evolution in BAM is H.

Figure~\ref{fig:loreneVsgrid} shows the $l=m=2$ waveforms aligned
before merger on the time window $t/M\in[250,739]$ (support of the black line.)
The waveforms are very similar; phase differences (black line) are below
$\Delta\phi\lesssim\pm0.2$~rad. This uncertainty is of the same order
magnitude of a conservative error bar estimated from convergence
tests. On the other hand, the HMNSs collapse within $150$~M ($2$~ms)
of each other. 

We conclude that the results consistently approach a continuum
limit when smaller grid spacings and sufficiently large boxes are
employed. Results from different initial data are also
consistent. However, care should be taken considering HMNS simulations
of several milliseconds since relatively small mass violations can lead
to quantitatively different behaviors. 
We have tested different grid setups, grid resolutions, and
independent initial data (when possible). Based on these results we
expect an uncertainty on the HMNS lifetime up to a maximum of 300M,
which is considerably shorter than the difference between model
$\Gamma_{000}$ and $\Gamma_{050}^{++}$.
Although not commonly used in
numerical relativity, a conservative AMR~\cite{Berger:1989} is
desirable; see~\cite{East:2011aa,Reisswig:2012nc} for the first
recent applications in the field.


\begin{thebibliography}{101}
\expandafter\ifx\csname natexlab\endcsname\relax\def\natexlab#1{#1}\fi
\expandafter\ifx\csname bibnamefont\endcsname\relax
  \def\bibnamefont#1{#1}\fi
\expandafter\ifx\csname bibfnamefont\endcsname\relax
  \def\bibfnamefont#1{#1}\fi
\expandafter\ifx\csname citenamefont\endcsname\relax
  \def\citenamefont#1{#1}\fi
\expandafter\ifx\csname url\endcsname\relax
  \def\url#1{\texttt{#1}}\fi
\expandafter\ifx\csname urlprefix\endcsname\relax\def\urlprefix{URL }\fi
\providecommand{\bibinfo}[2]{#2}
\providecommand{\eprint}[2][]{\url{#2}}

\bibitem[{\citenamefont{Lorimer}(2008)}]{Lorimer:2008se}
\bibinfo{author}{\bibfnamefont{D.~R.} \bibnamefont{Lorimer}},
  \bibinfo{journal}{Living Rev. Rel.} \textbf{\bibinfo{volume}{11}},
  \bibinfo{pages}{8} (\bibinfo{year}{2008}), \eprint{0811.0762}.

\bibitem[{\citenamefont{Lyne et~al.}(2004)\citenamefont{Lyne, Burgay, Kramer,
  Possenti, Manchester et~al.}}]{Lyne:2004cj}
\bibinfo{author}{\bibfnamefont{A.}~\bibnamefont{Lyne}},
  \bibinfo{author}{\bibfnamefont{M.}~\bibnamefont{Burgay}},
  \bibinfo{author}{\bibfnamefont{M.}~\bibnamefont{Kramer}},
  \bibinfo{author}{\bibfnamefont{A.}~\bibnamefont{Possenti}},
  \bibinfo{author}{\bibfnamefont{R.}~\bibnamefont{Manchester}},
  \bibnamefont{et~al.}, \bibinfo{journal}{Science}
  \textbf{\bibinfo{volume}{303}}, \bibinfo{pages}{1153} (\bibinfo{year}{2004}),
  \eprint{astro-ph/0401086}.

\bibitem[{\citenamefont{Burgay et~al.}(2003)\citenamefont{Burgay, D'Amico,
  Possenti, Manchester, Lyne et~al.}}]{Burgay:2003jj}
\bibinfo{author}{\bibfnamefont{M.}~\bibnamefont{Burgay}},
  \bibinfo{author}{\bibfnamefont{N.}~\bibnamefont{D'Amico}},
  \bibinfo{author}{\bibfnamefont{A.}~\bibnamefont{Possenti}},
  \bibinfo{author}{\bibfnamefont{R.}~\bibnamefont{Manchester}},
  \bibinfo{author}{\bibfnamefont{A.}~\bibnamefont{Lyne}}, \bibnamefont{et~al.},
  \bibinfo{journal}{Nature} \textbf{\bibinfo{volume}{426}},
  \bibinfo{pages}{531} (\bibinfo{year}{2003}), \eprint{astro-ph/0312071}.

\bibitem[{\citenamefont{Tichy}(2011)}]{Tichy:2011gw}
\bibinfo{author}{\bibfnamefont{W.}~\bibnamefont{Tichy}},
  \bibinfo{journal}{Phys.Rev.} \textbf{\bibinfo{volume}{D84}},
  \bibinfo{pages}{024041} (\bibinfo{year}{2011}), \eprint{1107.1440}.

\bibitem[{\citenamefont{Damour et~al.}(2012{\natexlab{a}})\citenamefont{Damour,
  Nagar, and Villain}}]{Damour:2012yf}
\bibinfo{author}{\bibfnamefont{T.}~\bibnamefont{Damour}},
  \bibinfo{author}{\bibfnamefont{A.}~\bibnamefont{Nagar}}, \bibnamefont{and}
  \bibinfo{author}{\bibfnamefont{L.}~\bibnamefont{Villain}},
  \bibinfo{journal}{Phys.Rev.} \textbf{\bibinfo{volume}{D85}},
  \bibinfo{pages}{123007} (\bibinfo{year}{2012}{\natexlab{a}}),
  \eprint{1203.4352}.

\bibitem[{\citenamefont{Brown et~al.}(2012)\citenamefont{Brown, Harry,
  Lundgren, and Nitz}}]{Brown:2012qf}
\bibinfo{author}{\bibfnamefont{D.~A.} \bibnamefont{Brown}},
  \bibinfo{author}{\bibfnamefont{I.}~\bibnamefont{Harry}},
  \bibinfo{author}{\bibfnamefont{A.}~\bibnamefont{Lundgren}}, \bibnamefont{and}
  \bibinfo{author}{\bibfnamefont{A.~H.} \bibnamefont{Nitz}},
  \bibinfo{journal}{Phys.Rev.} \textbf{\bibinfo{volume}{D86}},
  \bibinfo{pages}{084017} (\bibinfo{year}{2012}), \eprint{1207.6406}.

\bibitem[{\citenamefont{Hannam et~al.}(2013)\citenamefont{Hannam, Brown,
  Fairhurst, Fryer, and Harry}}]{Hannam:2013uu}
\bibinfo{author}{\bibfnamefont{M.}~\bibnamefont{Hannam}},
  \bibinfo{author}{\bibfnamefont{D.~A.} \bibnamefont{Brown}},
  \bibinfo{author}{\bibfnamefont{S.}~\bibnamefont{Fairhurst}},
  \bibinfo{author}{\bibfnamefont{C.~L.} \bibnamefont{Fryer}}, \bibnamefont{and}
  \bibinfo{author}{\bibfnamefont{I.~W.} \bibnamefont{Harry}},
  \bibinfo{journal}{Astrophys.J.} \textbf{\bibinfo{volume}{766}},
  \bibinfo{pages}{L14} (\bibinfo{year}{2013}), \eprint{1301.5616}.

\bibitem[{\citenamefont{Lo and Lin}(2011)}]{Lo:2010bj}
\bibinfo{author}{\bibfnamefont{K.-W.} \bibnamefont{Lo}} \bibnamefont{and}
  \bibinfo{author}{\bibfnamefont{L.-M.} \bibnamefont{Lin}},
  \bibinfo{journal}{Astrophys.J.} \textbf{\bibinfo{volume}{728}},
  \bibinfo{pages}{12} (\bibinfo{year}{2011}), \eprint{1011.3563}.

\bibitem[{\citenamefont{Duez et~al.}(2004)\citenamefont{Duez, Liu, Shapiro, and
  Stephens}}]{Duez:2004nf}
\bibinfo{author}{\bibfnamefont{M.~D.} \bibnamefont{Duez}},
  \bibinfo{author}{\bibfnamefont{Y.~T.} \bibnamefont{Liu}},
  \bibinfo{author}{\bibfnamefont{S.~L.} \bibnamefont{Shapiro}},
  \bibnamefont{and} \bibinfo{author}{\bibfnamefont{B.~C.}
  \bibnamefont{Stephens}}, \bibinfo{journal}{Phys.Rev.}
  \textbf{\bibinfo{volume}{D69}}, \bibinfo{pages}{104030}
  (\bibinfo{year}{2004}), \eprint{astro-ph/0402502}.

\bibitem[{\citenamefont{Abadie et~al.}(2010)}]{Abadie:2010cf}
\bibinfo{author}{\bibfnamefont{J.}~\bibnamefont{Abadie}} \bibnamefont{et~al.}
  (\bibinfo{collaboration}{LIGO Scientific Collaboration, Virgo
  Collaboration}), \bibinfo{journal}{Class.Quant.Grav.}
  \textbf{\bibinfo{volume}{27}}, \bibinfo{pages}{173001}
  (\bibinfo{year}{2010}), \eprint{1003.2480}.

\bibitem[{\citenamefont{Aasi et~al.}(2013)}]{Aasi:2013wya}
\bibinfo{author}{\bibfnamefont{J.}~\bibnamefont{Aasi}} \bibnamefont{et~al.}
  (\bibinfo{collaboration}{LIGO Scientific Collaboration, Virgo Collaboration})
  (\bibinfo{year}{2013}), \eprint{1304.0670}.

\bibitem[{\citenamefont{Ajith}(2011)}]{Ajith:2011ec}
\bibinfo{author}{\bibfnamefont{P.}~\bibnamefont{Ajith}},
  \bibinfo{journal}{Phys.Rev.} \textbf{\bibinfo{volume}{D84}},
  \bibinfo{pages}{084037} (\bibinfo{year}{2011}), \eprint{1107.1267}.

\bibitem[{\citenamefont{Read et~al.}(2009)\citenamefont{Read, Markakis,
  Shibata, Uryu, Creighton et~al.}}]{Read:2009yp}
\bibinfo{author}{\bibfnamefont{J.~S.} \bibnamefont{Read}},
  \bibinfo{author}{\bibfnamefont{C.}~\bibnamefont{Markakis}},
  \bibinfo{author}{\bibfnamefont{M.}~\bibnamefont{Shibata}},
  \bibinfo{author}{\bibfnamefont{K.}~\bibnamefont{Uryu}},
  \bibinfo{author}{\bibfnamefont{J.~D.} \bibnamefont{Creighton}},
  \bibnamefont{et~al.}, \bibinfo{journal}{Phys.Rev.}
  \textbf{\bibinfo{volume}{D79}}, \bibinfo{pages}{124033}
  (\bibinfo{year}{2009}), \eprint{0901.3258}.

\bibitem[{\citenamefont{Del~Pozzo et~al.}(2013)\citenamefont{Del~Pozzo, Li,
  Agathos, Broeck, and Vitale}}]{DelPozzo:2013ala}
\bibinfo{author}{\bibfnamefont{W.}~\bibnamefont{Del~Pozzo}},
  \bibinfo{author}{\bibfnamefont{T.~G.~F.} \bibnamefont{Li}},
  \bibinfo{author}{\bibfnamefont{M.}~\bibnamefont{Agathos}},
  \bibinfo{author}{\bibfnamefont{C.~V.~D.} \bibnamefont{Broeck}},
  \bibnamefont{and} \bibinfo{author}{\bibfnamefont{S.}~\bibnamefont{Vitale}},
  \bibinfo{journal}{Phys.~Rev.~Lett.} \textbf{\bibinfo{volume}{111}},
  \bibinfo{pages}{071101} (\bibinfo{year}{2013}), \eprint{1307.8338}.

\bibitem[{\citenamefont{Baiotti et~al.}(2011)\citenamefont{Baiotti, Damour,
  Giacomazzo, Nagar, and Rezzolla}}]{Baiotti:2011am}
\bibinfo{author}{\bibfnamefont{L.}~\bibnamefont{Baiotti}},
  \bibinfo{author}{\bibfnamefont{T.}~\bibnamefont{Damour}},
  \bibinfo{author}{\bibfnamefont{B.}~\bibnamefont{Giacomazzo}},
  \bibinfo{author}{\bibfnamefont{A.}~\bibnamefont{Nagar}}, \bibnamefont{and}
  \bibinfo{author}{\bibfnamefont{L.}~\bibnamefont{Rezzolla}},
  \bibinfo{journal}{Phys. Rev.} \textbf{\bibinfo{volume}{D84}},
  \bibinfo{pages}{024017} (\bibinfo{year}{2011}), \eprint{1103.3874}.

\bibitem[{\citenamefont{Bernuzzi
  et~al.}(2012{\natexlab{a}})\citenamefont{Bernuzzi, Thierfelder, and
  Br{\"u}gmann}}]{Bernuzzi:2011aq}
\bibinfo{author}{\bibfnamefont{S.}~\bibnamefont{Bernuzzi}},
  \bibinfo{author}{\bibfnamefont{M.}~\bibnamefont{Thierfelder}},
  \bibnamefont{and}
  \bibinfo{author}{\bibfnamefont{B.}~\bibnamefont{Br{\"u}gmann}},
  \bibinfo{journal}{Phys.Rev.} \textbf{\bibinfo{volume}{D85}},
  \bibinfo{pages}{104030} (\bibinfo{year}{2012}{\natexlab{a}}),
  \eprint{1109.3611}.

\bibitem[{\citenamefont{Bernuzzi
  et~al.}(2012{\natexlab{b}})\citenamefont{Bernuzzi, Nagar, Thierfelder, and
  Br{\"u}gmann}}]{Bernuzzi:2012ci}
\bibinfo{author}{\bibfnamefont{S.}~\bibnamefont{Bernuzzi}},
  \bibinfo{author}{\bibfnamefont{A.}~\bibnamefont{Nagar}},
  \bibinfo{author}{\bibfnamefont{M.}~\bibnamefont{Thierfelder}},
  \bibnamefont{and}
  \bibinfo{author}{\bibfnamefont{B.}~\bibnamefont{Br{\"u}gmann}},
  \bibinfo{journal}{Phys.Rev.} \textbf{\bibinfo{volume}{D86}},
  \bibinfo{pages}{044030} (\bibinfo{year}{2012}{\natexlab{b}}),
  \eprint{1205.3403}.

\bibitem[{\citenamefont{Hotokezaka et~al.}(2013)\citenamefont{Hotokezaka,
  Kyutoku, and Shibata}}]{Hotokezaka:2013mm}
\bibinfo{author}{\bibfnamefont{K.}~\bibnamefont{Hotokezaka}},
  \bibinfo{author}{\bibfnamefont{K.}~\bibnamefont{Kyutoku}}, \bibnamefont{and}
  \bibinfo{author}{\bibfnamefont{M.}~\bibnamefont{Shibata}},
  \bibinfo{journal}{Phys.Rev.} \textbf{\bibinfo{volume}{D87}},
  \bibinfo{pages}{044001} (\bibinfo{year}{2013}), \eprint{1301.3555}.

\bibitem[{\citenamefont{Baumgarte et~al.}(1997)\citenamefont{Baumgarte, Cook,
  Scheel, Shapiro, and Teukolsky}}]{Baumgarte:1997xi}
\bibinfo{author}{\bibfnamefont{T.}~\bibnamefont{Baumgarte}},
  \bibinfo{author}{\bibfnamefont{G.}~\bibnamefont{Cook}},
  \bibinfo{author}{\bibfnamefont{M.}~\bibnamefont{Scheel}},
  \bibinfo{author}{\bibfnamefont{S.}~\bibnamefont{Shapiro}}, \bibnamefont{and}
  \bibinfo{author}{\bibfnamefont{S.}~\bibnamefont{Teukolsky}},
  \bibinfo{journal}{Phys.Rev.Lett.} \textbf{\bibinfo{volume}{79}},
  \bibinfo{pages}{1182} (\bibinfo{year}{1997}), \eprint{gr-qc/9704024}.

\bibitem[{\citenamefont{Mathews et~al.}(1998)\citenamefont{Mathews, Marronetti,
  and Wilson}}]{Mathews:1997pf}
\bibinfo{author}{\bibfnamefont{G.}~\bibnamefont{Mathews}},
  \bibinfo{author}{\bibfnamefont{P.}~\bibnamefont{Marronetti}},
  \bibnamefont{and} \bibinfo{author}{\bibfnamefont{J.}~\bibnamefont{Wilson}},
  \bibinfo{journal}{Phys.Rev.} \textbf{\bibinfo{volume}{D58}},
  \bibinfo{pages}{043003} (\bibinfo{year}{1998}), \eprint{gr-qc/9710140}.

\bibitem[{\citenamefont{Marronetti et~al.}(1998)\citenamefont{Marronetti,
  Mathews, and Wilson}}]{Marronetti:1998xv}
\bibinfo{author}{\bibfnamefont{P.}~\bibnamefont{Marronetti}},
  \bibinfo{author}{\bibfnamefont{G.}~\bibnamefont{Mathews}}, \bibnamefont{and}
  \bibinfo{author}{\bibfnamefont{J.}~\bibnamefont{Wilson}},
  \bibinfo{journal}{Phys.Rev.} \textbf{\bibinfo{volume}{D58}},
  \bibinfo{pages}{107503} (\bibinfo{year}{1998}), \eprint{gr-qc/9803093}.

\bibitem[{\citenamefont{Bonazzola et~al.}(1999)\citenamefont{Bonazzola,
  Gourgoulhon, and Marck}}]{Bonazzola:1998yq}
\bibinfo{author}{\bibfnamefont{S.}~\bibnamefont{Bonazzola}},
  \bibinfo{author}{\bibfnamefont{E.}~\bibnamefont{Gourgoulhon}},
  \bibnamefont{and} \bibinfo{author}{\bibfnamefont{J.-A.} \bibnamefont{Marck}},
  \bibinfo{journal}{Phys.Rev.Lett.} \textbf{\bibinfo{volume}{82}},
  \bibinfo{pages}{892} (\bibinfo{year}{1999}), \eprint{gr-qc/9810072}.

\bibitem[{\citenamefont{Marronetti et~al.}(1999)\citenamefont{Marronetti,
  Mathews, and Wilson}}]{Marronetti:1999ya}
\bibinfo{author}{\bibfnamefont{P.}~\bibnamefont{Marronetti}},
  \bibinfo{author}{\bibfnamefont{G.}~\bibnamefont{Mathews}}, \bibnamefont{and}
  \bibinfo{author}{\bibfnamefont{J.}~\bibnamefont{Wilson}},
  \bibinfo{journal}{Phys.Rev.} \textbf{\bibinfo{volume}{D60}},
  \bibinfo{pages}{087301} (\bibinfo{year}{1999}), \eprint{gr-qc/9906088}.

\bibitem[{\citenamefont{Uryu and Eriguchi}(2000)}]{Uryu:1999uu}
\bibinfo{author}{\bibfnamefont{K.}~\bibnamefont{Uryu}} \bibnamefont{and}
  \bibinfo{author}{\bibfnamefont{Y.}~\bibnamefont{Eriguchi}},
  \bibinfo{journal}{Phys.Rev.} \textbf{\bibinfo{volume}{D61}},
  \bibinfo{pages}{124023} (\bibinfo{year}{2000}), \eprint{gr-qc/9908059}.

\bibitem[{\citenamefont{Taniguchi and Gourgoulhon}(2002)}]{Taniguchi:2002ns}
\bibinfo{author}{\bibfnamefont{K.}~\bibnamefont{Taniguchi}} \bibnamefont{and}
  \bibinfo{author}{\bibfnamefont{E.}~\bibnamefont{Gourgoulhon}},
  \bibinfo{journal}{Phys. Rev.} \textbf{\bibinfo{volume}{D66}},
  \bibinfo{pages}{104019} (\bibinfo{year}{2002}), \eprint{gr-qc/0207098}.

\bibitem[{\citenamefont{Uryu et~al.}(2006)\citenamefont{Uryu, Limousin,
  Friedman, Gourgoulhon, and Shibata}}]{Uryu:2005vv}
\bibinfo{author}{\bibfnamefont{K.}~\bibnamefont{Uryu}},
  \bibinfo{author}{\bibfnamefont{F.}~\bibnamefont{Limousin}},
  \bibinfo{author}{\bibfnamefont{J.~L.} \bibnamefont{Friedman}},
  \bibinfo{author}{\bibfnamefont{E.}~\bibnamefont{Gourgoulhon}},
  \bibnamefont{and} \bibinfo{author}{\bibfnamefont{M.}~\bibnamefont{Shibata}},
  \bibinfo{journal}{Phys.Rev.Lett.} \textbf{\bibinfo{volume}{97}},
  \bibinfo{pages}{171101} (\bibinfo{year}{2006}), \eprint{gr-qc/0511136}.

\bibitem[{\citenamefont{Uryu et~al.}(2009)\citenamefont{Uryu, Limousin,
  Friedman, Gourgoulhon, and Shibata}}]{Uryu:2009ye}
\bibinfo{author}{\bibfnamefont{K.}~\bibnamefont{Uryu}},
  \bibinfo{author}{\bibfnamefont{F.}~\bibnamefont{Limousin}},
  \bibinfo{author}{\bibfnamefont{J.~L.} \bibnamefont{Friedman}},
  \bibinfo{author}{\bibfnamefont{E.}~\bibnamefont{Gourgoulhon}},
  \bibnamefont{and} \bibinfo{author}{\bibfnamefont{M.}~\bibnamefont{Shibata}},
  \bibinfo{journal}{Phys. Rev.} \textbf{\bibinfo{volume}{D80}},
  \bibinfo{pages}{124004} (\bibinfo{year}{2009}), \eprint{0908.0579}.

\bibitem[{\citenamefont{{Bildsten} and
  {Cutler}}(1992)}]{Bildsten:1992ApJ...400..175B}
\bibinfo{author}{\bibfnamefont{L.}~\bibnamefont{{Bildsten}}} \bibnamefont{and}
  \bibinfo{author}{\bibfnamefont{C.}~\bibnamefont{{Cutler}}},
  \bibinfo{journal}{Astrophys. J.} \textbf{\bibinfo{volume}{400}},
  \bibinfo{pages}{175} (\bibinfo{year}{1992}).

\bibitem[{\citenamefont{Sekiguchi
  et~al.}(2011{\natexlab{a}})\citenamefont{Sekiguchi, Kiuchi, Kyutoku, and
  Shibata}}]{Sekiguchi:2011zd}
\bibinfo{author}{\bibfnamefont{Y.}~\bibnamefont{Sekiguchi}},
  \bibinfo{author}{\bibfnamefont{K.}~\bibnamefont{Kiuchi}},
  \bibinfo{author}{\bibfnamefont{K.}~\bibnamefont{Kyutoku}}, \bibnamefont{and}
  \bibinfo{author}{\bibfnamefont{M.}~\bibnamefont{Shibata}},
  \bibinfo{journal}{Phys.Rev.Lett.} \textbf{\bibinfo{volume}{107}},
  \bibinfo{pages}{051102} (\bibinfo{year}{2011}{\natexlab{a}}),
  \eprint{1105.2125}.

\bibitem[{\citenamefont{Sekiguchi
  et~al.}(2011{\natexlab{b}})\citenamefont{Sekiguchi, Kiuchi, Kyutoku, and
  Shibata}}]{Sekiguchi:2011mc}
\bibinfo{author}{\bibfnamefont{Y.}~\bibnamefont{Sekiguchi}},
  \bibinfo{author}{\bibfnamefont{K.}~\bibnamefont{Kiuchi}},
  \bibinfo{author}{\bibfnamefont{K.}~\bibnamefont{Kyutoku}}, \bibnamefont{and}
  \bibinfo{author}{\bibfnamefont{M.}~\bibnamefont{Shibata}},
  \bibinfo{journal}{Phys.Rev.Lett.} \textbf{\bibinfo{volume}{107}},
  \bibinfo{pages}{211101} (\bibinfo{year}{2011}{\natexlab{b}}),
  \eprint{1110.4442}.

\bibitem[{\citenamefont{Palenzuela et~al.}(2013)\citenamefont{Palenzuela,
  Lehner, Ponce, Liebling, Anderson et~al.}}]{Palenzuela:2013hu}
\bibinfo{author}{\bibfnamefont{C.}~\bibnamefont{Palenzuela}},
  \bibinfo{author}{\bibfnamefont{L.}~\bibnamefont{Lehner}},
  \bibinfo{author}{\bibfnamefont{M.}~\bibnamefont{Ponce}},
  \bibinfo{author}{\bibfnamefont{S.~L.} \bibnamefont{Liebling}},
  \bibinfo{author}{\bibfnamefont{M.}~\bibnamefont{Anderson}},
  \bibnamefont{et~al.}, \bibinfo{journal}{Phys.Rev.Lett.}
  \textbf{\bibinfo{volume}{111}}, \bibinfo{pages}{061105}
  (\bibinfo{year}{2013}), \eprint{1301.7074}.

\bibitem[{\citenamefont{Gold et~al.}(2012)\citenamefont{Gold, Bernuzzi,
  Thierfelder, Br{\"u}gmann, and Pretorius}}]{Gold:2011df}
\bibinfo{author}{\bibfnamefont{R.}~\bibnamefont{Gold}},
  \bibinfo{author}{\bibfnamefont{S.}~\bibnamefont{Bernuzzi}},
  \bibinfo{author}{\bibfnamefont{M.}~\bibnamefont{Thierfelder}},
  \bibinfo{author}{\bibfnamefont{B.}~\bibnamefont{Br{\"u}gmann}},
  \bibnamefont{and}
  \bibinfo{author}{\bibfnamefont{F.}~\bibnamefont{Pretorius}},
  \bibinfo{journal}{Phys.Rev.} \textbf{\bibinfo{volume}{D86}},
  \bibinfo{pages}{121501} (\bibinfo{year}{2012}), \eprint{1109.5128}.

\bibitem[{\citenamefont{East and Pretorius}(2012)}]{East:2012ww}
\bibinfo{author}{\bibfnamefont{W.~E.} \bibnamefont{East}} \bibnamefont{and}
  \bibinfo{author}{\bibfnamefont{F.}~\bibnamefont{Pretorius}},
  \bibinfo{journal}{Astrophys.J.} \textbf{\bibinfo{volume}{760}},
  \bibinfo{pages}{L4} (\bibinfo{year}{2012}), \eprint{1208.5279}.

\bibitem[{\citenamefont{Faber and Rasio}(2012)}]{Faber:2012rw}
\bibinfo{author}{\bibfnamefont{J.~A.} \bibnamefont{Faber}} \bibnamefont{and}
  \bibinfo{author}{\bibfnamefont{F.~A.} \bibnamefont{Rasio}},
  \bibinfo{journal}{Living Rev.Rel.} \textbf{\bibinfo{volume}{15}},
  \bibinfo{pages}{8} (\bibinfo{year}{2012}), \eprint{1204.3858}.

\bibitem[{\citenamefont{Tichy}(2012)}]{Tichy:2012rp}
\bibinfo{author}{\bibfnamefont{W.}~\bibnamefont{Tichy}},
  \bibinfo{journal}{Phys. Rev. D} \textbf{\bibinfo{volume}{86}},
  \bibinfo{pages}{064024} (\bibinfo{year}{2012}), \eprint{1209.5336}.

\bibitem[{\citenamefont{Marronetti and Shapiro}(2003)}]{Marronetti:2003gk}
\bibinfo{author}{\bibfnamefont{P.}~\bibnamefont{Marronetti}} \bibnamefont{and}
  \bibinfo{author}{\bibfnamefont{S.~L.} \bibnamefont{Shapiro}},
  \bibinfo{journal}{Phys.Rev.} \textbf{\bibinfo{volume}{D68}},
  \bibinfo{pages}{104024} (\bibinfo{year}{2003}), \eprint{gr-qc/0306075}.

\bibitem[{\citenamefont{Baumgarte and Shapiro}(2009)}]{Baumgarte:2009fw}
\bibinfo{author}{\bibfnamefont{T.~W.} \bibnamefont{Baumgarte}}
  \bibnamefont{and} \bibinfo{author}{\bibfnamefont{S.~L.}
  \bibnamefont{Shapiro}}, \bibinfo{journal}{Phys.Rev.}
  \textbf{\bibinfo{volume}{D80}}, \bibinfo{pages}{064009}
  (\bibinfo{year}{2009}), \eprint{0909.0952}.

\bibitem[{\citenamefont{Damour et~al.}(2012{\natexlab{b}})\citenamefont{Damour,
  Nagar, Pollney, and Reisswig}}]{Damour:2011fu}
\bibinfo{author}{\bibfnamefont{T.}~\bibnamefont{Damour}},
  \bibinfo{author}{\bibfnamefont{A.}~\bibnamefont{Nagar}},
  \bibinfo{author}{\bibfnamefont{D.}~\bibnamefont{Pollney}}, \bibnamefont{and}
  \bibinfo{author}{\bibfnamefont{C.}~\bibnamefont{Reisswig}},
  \bibinfo{journal}{Phys.Rev.Lett.} \textbf{\bibinfo{volume}{108}},
  \bibinfo{pages}{131101} (\bibinfo{year}{2012}{\natexlab{b}}),
  \eprint{1110.2938}.

\bibitem[{\citenamefont{Shibata and Uryu}(2000)}]{Shibata:1999wm}
\bibinfo{author}{\bibfnamefont{M.}~\bibnamefont{Shibata}} \bibnamefont{and}
  \bibinfo{author}{\bibfnamefont{K.}~\bibnamefont{Uryu}},
  \bibinfo{journal}{Phys. Rev.} \textbf{\bibinfo{volume}{D61}},
  \bibinfo{pages}{064001} (\bibinfo{year}{2000}), \eprint{gr-qc/9911058}.

\bibitem[{\citenamefont{Oechslin and Janka}(2007)}]{Oechslin:2007gn}
\bibinfo{author}{\bibfnamefont{R.}~\bibnamefont{Oechslin}} \bibnamefont{and}
  \bibinfo{author}{\bibfnamefont{H.~T.} \bibnamefont{Janka}},
  \bibinfo{journal}{Phys. Rev. Lett.} \textbf{\bibinfo{volume}{99}},
  \bibinfo{pages}{121102} (\bibinfo{year}{2007}), \eprint{astro-ph/0702228}.

\bibitem[{\citenamefont{Kastaun et~al.}(2013)\citenamefont{Kastaun, Galeazzi,
  Alic, Rezzolla, and Font}}]{Kastaun:2013mv}
\bibinfo{author}{\bibfnamefont{W.}~\bibnamefont{Kastaun}},
  \bibinfo{author}{\bibfnamefont{F.}~\bibnamefont{Galeazzi}},
  \bibinfo{author}{\bibfnamefont{D.}~\bibnamefont{Alic}},
  \bibinfo{author}{\bibfnamefont{L.}~\bibnamefont{Rezzolla}}, \bibnamefont{and}
  \bibinfo{author}{\bibfnamefont{J.~A.} \bibnamefont{Font}},
  \bibinfo{journal}{Phys.Rev.} \textbf{\bibinfo{volume}{D88}},
  \bibinfo{pages}{021501} (\bibinfo{year}{2013}), \eprint{1301.7348}.

\bibitem[{\citenamefont{Tsatsin and Marronetti}(2013)}]{Tsatsin:2013jca}
\bibinfo{author}{\bibfnamefont{P.}~\bibnamefont{Tsatsin}} \bibnamefont{and}
  \bibinfo{author}{\bibfnamefont{P.}~\bibnamefont{Marronetti}},
  \bibinfo{journal}{Phys.Rev.} \textbf{\bibinfo{volume}{D88}},
  \bibinfo{pages}{064060} (\bibinfo{year}{2013}), \eprint{1303.6692}.

\bibitem[{\citenamefont{Wilson and Mathews}(1995)}]{Wilson:1995uh}
\bibinfo{author}{\bibfnamefont{J.}~\bibnamefont{Wilson}} \bibnamefont{and}
  \bibinfo{author}{\bibfnamefont{G.}~\bibnamefont{Mathews}},
  \bibinfo{journal}{Phys.Rev.Lett.} \textbf{\bibinfo{volume}{75}},
  \bibinfo{pages}{4161} (\bibinfo{year}{1995}).

\bibitem[{\citenamefont{Wilson et~al.}(1996)\citenamefont{Wilson, Mathews, and
  Marronetti}}]{Wilson:1996ty}
\bibinfo{author}{\bibfnamefont{J.}~\bibnamefont{Wilson}},
  \bibinfo{author}{\bibfnamefont{G.}~\bibnamefont{Mathews}}, \bibnamefont{and}
  \bibinfo{author}{\bibfnamefont{P.}~\bibnamefont{Marronetti}},
  \bibinfo{journal}{Phys.Rev.} \textbf{\bibinfo{volume}{D54}},
  \bibinfo{pages}{1317} (\bibinfo{year}{1996}), \eprint{gr-qc/9601017}.

\bibitem[{\citenamefont{York}(1999)}]{York:1998hy}
\bibinfo{author}{\bibfnamefont{J.}~\bibnamefont{York},
  \bibfnamefont{James~W.}}, \bibinfo{journal}{Phys.Rev.Lett.}
  \textbf{\bibinfo{volume}{82}}, \bibinfo{pages}{1350} (\bibinfo{year}{1999}),
  \eprint{gr-qc/9810051}.

\bibitem[{\citenamefont{Tichy}(2006)}]{Tichy:2006qn}
\bibinfo{author}{\bibfnamefont{W.}~\bibnamefont{Tichy}},
  \bibinfo{journal}{Phys.Rev.} \textbf{\bibinfo{volume}{D74}},
  \bibinfo{pages}{084005} (\bibinfo{year}{2006}), \eprint{gr-qc/0609087}.

\bibitem[{\citenamefont{Tichy}(2009{\natexlab{a}})}]{Tichy:2009yr}
\bibinfo{author}{\bibfnamefont{W.}~\bibnamefont{Tichy}},
  \bibinfo{journal}{Class.Quant.Grav.} \textbf{\bibinfo{volume}{26}},
  \bibinfo{pages}{175018} (\bibinfo{year}{2009}{\natexlab{a}}),
  \eprint{0908.0620}.

\bibitem[{\citenamefont{Tichy}(2009{\natexlab{b}})}]{Tichy:2009zr}
\bibinfo{author}{\bibfnamefont{W.}~\bibnamefont{Tichy}},
  \bibinfo{journal}{Phys.Rev.} \textbf{\bibinfo{volume}{D80}},
  \bibinfo{pages}{104034} (\bibinfo{year}{2009}{\natexlab{b}}),
  \eprint{0911.0973}.

\bibitem[{\citenamefont{Walther et~al.}(2009)\citenamefont{Walther,
  Br{\"u}gmann, and M{\"u}ller}}]{Walther:2009ng}
\bibinfo{author}{\bibfnamefont{B.}~\bibnamefont{Walther}},
  \bibinfo{author}{\bibfnamefont{B.}~\bibnamefont{Br{\"u}gmann}},
  \bibnamefont{and}
  \bibinfo{author}{\bibfnamefont{D.}~\bibnamefont{M{\"u}ller}},
  \bibinfo{journal}{Phys.Rev.} \textbf{\bibinfo{volume}{D79}},
  \bibinfo{pages}{124040} (\bibinfo{year}{2009}), \eprint{0901.0993}.

\bibitem[{\citenamefont{Thierfelder
  et~al.}(2011{\natexlab{a}})\citenamefont{Thierfelder, Bernuzzi, and
  Br{\"u}gmann}}]{Thierfelder:2011yi}
\bibinfo{author}{\bibfnamefont{M.}~\bibnamefont{Thierfelder}},
  \bibinfo{author}{\bibfnamefont{S.}~\bibnamefont{Bernuzzi}}, \bibnamefont{and}
  \bibinfo{author}{\bibfnamefont{B.}~\bibnamefont{Br{\"u}gmann}},
  \bibinfo{journal}{Phys.Rev.} \textbf{\bibinfo{volume}{D84}},
  \bibinfo{pages}{044012} (\bibinfo{year}{2011}{\natexlab{a}}),
  \eprint{1104.4751}.

\bibitem[{\citenamefont{Br{\"u}gmann et~al.}(2008)\citenamefont{Br{\"u}gmann,
  Gonzalez, Hannam, Husa, Sperhake et~al.}}]{Brugmann:2008zz}
\bibinfo{author}{\bibfnamefont{B.}~\bibnamefont{Br{\"u}gmann}},
  \bibinfo{author}{\bibfnamefont{J.~A.} \bibnamefont{Gonzalez}},
  \bibinfo{author}{\bibfnamefont{M.}~\bibnamefont{Hannam}},
  \bibinfo{author}{\bibfnamefont{S.}~\bibnamefont{Husa}},
  \bibinfo{author}{\bibfnamefont{U.}~\bibnamefont{Sperhake}},
  \bibnamefont{et~al.}, \bibinfo{journal}{Phys.Rev.}
  \textbf{\bibinfo{volume}{D77}}, \bibinfo{pages}{024027}
  (\bibinfo{year}{2008}), \eprint{gr-qc/0610128}.

\bibitem[{\citenamefont{Br{\"u}gmann et~al.}(2004)\citenamefont{Br{\"u}gmann,
  Tichy, and Jansen}}]{Bruegmann:2003aw}
\bibinfo{author}{\bibfnamefont{B.}~\bibnamefont{Br{\"u}gmann}},
  \bibinfo{author}{\bibfnamefont{W.}~\bibnamefont{Tichy}}, \bibnamefont{and}
  \bibinfo{author}{\bibfnamefont{N.}~\bibnamefont{Jansen}},
  \bibinfo{journal}{Phys. Rev. Lett.} \textbf{\bibinfo{volume}{92}},
  \bibinfo{pages}{211101} (\bibinfo{year}{2004}), \eprint{gr-qc/0312112}.

\bibitem[{\citenamefont{Br{\"u}gmann}(1999)}]{Bruegmann:1997uc}
\bibinfo{author}{\bibfnamefont{B.}~\bibnamefont{Br{\"u}gmann}},
  \bibinfo{journal}{Int. J. Mod. Phys.} \textbf{\bibinfo{volume}{D8}},
  \bibinfo{pages}{85} (\bibinfo{year}{1999}), \eprint{gr-qc/9708035}.

\bibitem[{\citenamefont{Nakamura et~al.}(1987)\citenamefont{Nakamura, Oohara,
  and Kojima}}]{Nakamura:1987zz}
\bibinfo{author}{\bibfnamefont{T.}~\bibnamefont{Nakamura}},
  \bibinfo{author}{\bibfnamefont{K.}~\bibnamefont{Oohara}}, \bibnamefont{and}
  \bibinfo{author}{\bibfnamefont{Y.}~\bibnamefont{Kojima}},
  \bibinfo{journal}{Prog. Theor. Phys. Suppl.} \textbf{\bibinfo{volume}{90}},
  \bibinfo{pages}{1} (\bibinfo{year}{1987}).

\bibitem[{\citenamefont{Shibata and Nakamura}(1995)}]{Shibata:1995we}
\bibinfo{author}{\bibfnamefont{M.}~\bibnamefont{Shibata}} \bibnamefont{and}
  \bibinfo{author}{\bibfnamefont{T.}~\bibnamefont{Nakamura}},
  \bibinfo{journal}{Phys. Rev.} \textbf{\bibinfo{volume}{D52}},
  \bibinfo{pages}{5428} (\bibinfo{year}{1995}).

\bibitem[{\citenamefont{Baumgarte and Shapiro}(1999)}]{Baumgarte:1998te}
\bibinfo{author}{\bibfnamefont{T.~W.} \bibnamefont{Baumgarte}}
  \bibnamefont{and} \bibinfo{author}{\bibfnamefont{S.~L.}
  \bibnamefont{Shapiro}}, \bibinfo{journal}{Phys. Rev.}
  \textbf{\bibinfo{volume}{D59}}, \bibinfo{pages}{024007}
  (\bibinfo{year}{1999}), \eprint{gr-qc/9810065}.

\bibitem[{\citenamefont{Bona et~al.}(1996)\citenamefont{Bona, Mass{\'o}, Stela,
  and Seidel}}]{Bona:1994a}
\bibinfo{author}{\bibfnamefont{C.}~\bibnamefont{Bona}},
  \bibinfo{author}{\bibfnamefont{J.}~\bibnamefont{Mass{\'o}}},
  \bibinfo{author}{\bibfnamefont{J.}~\bibnamefont{Stela}}, \bibnamefont{and}
  \bibinfo{author}{\bibfnamefont{E.}~\bibnamefont{Seidel}}, in
  \emph{\bibinfo{booktitle}{The Seventh {M}arcel {G}rossmann Meeting: On Recent
  Developments in Theoretical and Experimental General Relativity, Gravitation,
  and Relativistic Field Theories}}, edited by
  \bibinfo{editor}{\bibfnamefont{R.~T.} \bibnamefont{Jantzen}},
  \bibinfo{editor}{\bibfnamefont{G.~M.} \bibnamefont{Keiser}},
  \bibnamefont{and} \bibinfo{editor}{\bibfnamefont{R.}~\bibnamefont{Ruffini}}
  (\bibinfo{publisher}{World {S}cientific}, \bibinfo{address}{Singapore},
  \bibinfo{year}{1996}).

\bibitem[{\citenamefont{Alcubierre et~al.}(2003)\citenamefont{Alcubierre,
  Br{\"u}gmann, Diener, Koppitz, Pollney et~al.}}]{Alcubierre:2002kk}
\bibinfo{author}{\bibfnamefont{M.}~\bibnamefont{Alcubierre}},
  \bibinfo{author}{\bibfnamefont{B.}~\bibnamefont{Br{\"u}gmann}},
  \bibinfo{author}{\bibfnamefont{P.}~\bibnamefont{Diener}},
  \bibinfo{author}{\bibfnamefont{M.}~\bibnamefont{Koppitz}},
  \bibinfo{author}{\bibfnamefont{D.}~\bibnamefont{Pollney}},
  \bibnamefont{et~al.}, \bibinfo{journal}{Phys.Rev.}
  \textbf{\bibinfo{volume}{D67}}, \bibinfo{pages}{084023}
  (\bibinfo{year}{2003}), \eprint{gr-qc/0206072}.

\bibitem[{\citenamefont{van Meter et~al.}(2006)\citenamefont{van Meter, Baker,
  Koppitz, and Choi}}]{vanMeter:2006vi}
\bibinfo{author}{\bibfnamefont{J.~R.} \bibnamefont{van Meter}},
  \bibinfo{author}{\bibfnamefont{J.~G.} \bibnamefont{Baker}},
  \bibinfo{author}{\bibfnamefont{M.}~\bibnamefont{Koppitz}}, \bibnamefont{and}
  \bibinfo{author}{\bibfnamefont{D.-I.} \bibnamefont{Choi}},
  \bibinfo{journal}{Phys. Rev.} \textbf{\bibinfo{volume}{D73}},
  \bibinfo{pages}{124011} (\bibinfo{year}{2006}), \eprint{gr-qc/0605030}.

\bibitem[{\citenamefont{Thierfelder
  et~al.}(2011{\natexlab{b}})\citenamefont{Thierfelder, Bernuzzi, Hilditch,
  Br{\"u}gmann, and Rezzolla}}]{Thierfelder:2010dv}
\bibinfo{author}{\bibfnamefont{M.}~\bibnamefont{Thierfelder}},
  \bibinfo{author}{\bibfnamefont{S.}~\bibnamefont{Bernuzzi}},
  \bibinfo{author}{\bibfnamefont{D.}~\bibnamefont{Hilditch}},
  \bibinfo{author}{\bibfnamefont{B.}~\bibnamefont{Br{\"u}gmann}},
  \bibnamefont{and} \bibinfo{author}{\bibfnamefont{L.}~\bibnamefont{Rezzolla}},
  \bibinfo{journal}{Phys.Rev.} \textbf{\bibinfo{volume}{D83}},
  \bibinfo{pages}{064022} (\bibinfo{year}{2011}{\natexlab{b}}),
  \eprint{1012.3703}.

\bibitem[{\citenamefont{Borges et~al.}(2008)\citenamefont{Borges, Carmona,
  Costa, and Don}}]{Borges20083191}
\bibinfo{author}{\bibfnamefont{R.}~\bibnamefont{Borges}},
  \bibinfo{author}{\bibfnamefont{M.}~\bibnamefont{Carmona}},
  \bibinfo{author}{\bibfnamefont{B.}~\bibnamefont{Costa}}, \bibnamefont{and}
  \bibinfo{author}{\bibfnamefont{W.~S.} \bibnamefont{Don}},
  \bibinfo{journal}{Journal of Computational Physics}
  \textbf{\bibinfo{volume}{227}}, \bibinfo{pages}{3191} (\bibinfo{year}{2008}).

\bibitem[{\citenamefont{Berger and Oliger}(1984)}]{Berger:1984zza}
\bibinfo{author}{\bibfnamefont{M.~J.} \bibnamefont{Berger}} \bibnamefont{and}
  \bibinfo{author}{\bibfnamefont{J.}~\bibnamefont{Oliger}},
  \bibinfo{journal}{J.Comput.Phys.} \textbf{\bibinfo{volume}{53}},
  \bibinfo{pages}{484} (\bibinfo{year}{1984}).

\bibitem[{\citenamefont{{Reisswig} and
  {Pollney}}(2011)}]{Reisswig:2011CQGra..28s5015R}
\bibinfo{author}{\bibfnamefont{C.}~\bibnamefont{{Reisswig}}} \bibnamefont{and}
  \bibinfo{author}{\bibfnamefont{D.}~\bibnamefont{{Pollney}}},
  \bibinfo{journal}{Classical and Quantum Gravity}
  \textbf{\bibinfo{volume}{28}}, \bibinfo{eid}{195015} (\bibinfo{year}{2011}),
  \eprint{1006.1632}.

\bibitem[{\citenamefont{Damour}(2001)}]{Damour:2001tu}
\bibinfo{author}{\bibfnamefont{T.}~\bibnamefont{Damour}},
  \bibinfo{journal}{Phys. Rev.} \textbf{\bibinfo{volume}{D64}},
  \bibinfo{pages}{124013} (\bibinfo{year}{2001}), \eprint{gr-qc/0103018}.

\bibitem[{\citenamefont{Campanelli et~al.}(2006)\citenamefont{Campanelli,
  Lousto, and Zlochower}}]{Campanelli:2006uy}
\bibinfo{author}{\bibfnamefont{M.}~\bibnamefont{Campanelli}},
  \bibinfo{author}{\bibfnamefont{C.}~\bibnamefont{Lousto}}, \bibnamefont{and}
  \bibinfo{author}{\bibfnamefont{Y.}~\bibnamefont{Zlochower}},
  \bibinfo{journal}{Phys.Rev.} \textbf{\bibinfo{volume}{D74}},
  \bibinfo{pages}{041501} (\bibinfo{year}{2006}), \eprint{gr-qc/0604012}.

\bibitem[{\citenamefont{Buonanno and Damour}(1999)}]{Buonanno:1998gg}
\bibinfo{author}{\bibfnamefont{A.}~\bibnamefont{Buonanno}} \bibnamefont{and}
  \bibinfo{author}{\bibfnamefont{T.}~\bibnamefont{Damour}},
  \bibinfo{journal}{Phys. Rev.} \textbf{\bibinfo{volume}{D59}},
  \bibinfo{pages}{084006} (\bibinfo{year}{1999}), \eprint{gr-qc/9811091}.

\bibitem[{\citenamefont{Buonanno and Damour}(2000)}]{Buonanno:2000ef}
\bibinfo{author}{\bibfnamefont{A.}~\bibnamefont{Buonanno}} \bibnamefont{and}
  \bibinfo{author}{\bibfnamefont{T.}~\bibnamefont{Damour}},
  \bibinfo{journal}{Phys. Rev.} \textbf{\bibinfo{volume}{D62}},
  \bibinfo{pages}{064015} (\bibinfo{year}{2000}), \eprint{gr-qc/0001013}.

\bibitem[{\citenamefont{Nagar}(2011)}]{Nagar:2011fx}
\bibinfo{author}{\bibfnamefont{A.}~\bibnamefont{Nagar}},
  \bibinfo{journal}{Phys.Rev.} \textbf{\bibinfo{volume}{D84}},
  \bibinfo{pages}{084028} (\bibinfo{year}{2011}), \eprint{1106.4349}.

\bibitem[{\citenamefont{Kidder et~al.}(1993)\citenamefont{Kidder, Will, and
  Wiseman}}]{Kidder:1992fr}
\bibinfo{author}{\bibfnamefont{L.~E.} \bibnamefont{Kidder}},
  \bibinfo{author}{\bibfnamefont{C.~M.} \bibnamefont{Will}}, \bibnamefont{and}
  \bibinfo{author}{\bibfnamefont{A.~G.} \bibnamefont{Wiseman}},
  \bibinfo{journal}{Phys.Rev.} \textbf{\bibinfo{volume}{D47}},
  \bibinfo{pages}{4183} (\bibinfo{year}{1993}), \eprint{gr-qc/9211025}.

\bibitem[{\citenamefont{Kidder}(1995)}]{Kidder:1995zr}
\bibinfo{author}{\bibfnamefont{L.~E.} \bibnamefont{Kidder}},
  \bibinfo{journal}{Phys.Rev.} \textbf{\bibinfo{volume}{D52}},
  \bibinfo{pages}{821} (\bibinfo{year}{1995}), \eprint{gr-qc/9506022}.

\bibitem[{\citenamefont{Tagoshi et~al.}(2001)\citenamefont{Tagoshi, Ohashi, and
  Owen}}]{Tagoshi:2000zg}
\bibinfo{author}{\bibfnamefont{H.}~\bibnamefont{Tagoshi}},
  \bibinfo{author}{\bibfnamefont{A.}~\bibnamefont{Ohashi}}, \bibnamefont{and}
  \bibinfo{author}{\bibfnamefont{B.~J.} \bibnamefont{Owen}},
  \bibinfo{journal}{Phys.Rev.} \textbf{\bibinfo{volume}{D63}},
  \bibinfo{pages}{044006} (\bibinfo{year}{2001}), \eprint{gr-qc/0010014}.

\bibitem[{\citenamefont{Blanchet et~al.}(2004)\citenamefont{Blanchet, Damour,
  Esposito-Farese, and Iyer}}]{Blanchet:2004ek}
\bibinfo{author}{\bibfnamefont{L.}~\bibnamefont{Blanchet}},
  \bibinfo{author}{\bibfnamefont{T.}~\bibnamefont{Damour}},
  \bibinfo{author}{\bibfnamefont{G.}~\bibnamefont{Esposito-Farese}},
  \bibnamefont{and} \bibinfo{author}{\bibfnamefont{B.~R.} \bibnamefont{Iyer}},
  \bibinfo{journal}{Phys.Rev.Lett.} \textbf{\bibinfo{volume}{93}},
  \bibinfo{pages}{091101} (\bibinfo{year}{2004}), \eprint{gr-qc/0406012}.

\bibitem[{\citenamefont{Faye et~al.}(2006)\citenamefont{Faye, Blanchet, and
  Buonanno}}]{Faye:2006gx}
\bibinfo{author}{\bibfnamefont{G.}~\bibnamefont{Faye}},
  \bibinfo{author}{\bibfnamefont{L.}~\bibnamefont{Blanchet}}, \bibnamefont{and}
  \bibinfo{author}{\bibfnamefont{A.}~\bibnamefont{Buonanno}},
  \bibinfo{journal}{Phys.Rev.} \textbf{\bibinfo{volume}{D74}},
  \bibinfo{pages}{104033} (\bibinfo{year}{2006}), \eprint{gr-qc/0605139}.

\bibitem[{\citenamefont{Damour et~al.}(2008{\natexlab{a}})\citenamefont{Damour,
  Jaranowski, and Sch{\"a}fer}}]{Damour:2007nc}
\bibinfo{author}{\bibfnamefont{T.}~\bibnamefont{Damour}},
  \bibinfo{author}{\bibfnamefont{P.}~\bibnamefont{Jaranowski}},
  \bibnamefont{and}
  \bibinfo{author}{\bibfnamefont{G.}~\bibnamefont{Sch{\"a}fer}},
  \bibinfo{journal}{Phys. Rev.} \textbf{\bibinfo{volume}{D77}},
  \bibinfo{pages}{064032} (\bibinfo{year}{2008}{\natexlab{a}}),
  \eprint{0711.1048}.

\bibitem[{\citenamefont{Steinhoff et~al.}(2008)\citenamefont{Steinhoff, Hergt,
  and Sch{\"a}fer}}]{Steinhoff:2007mb}
\bibinfo{author}{\bibfnamefont{J.}~\bibnamefont{Steinhoff}},
  \bibinfo{author}{\bibfnamefont{S.}~\bibnamefont{Hergt}}, \bibnamefont{and}
  \bibinfo{author}{\bibfnamefont{G.}~\bibnamefont{Sch{\"a}fer}},
  \bibinfo{journal}{Phys.Rev.} \textbf{\bibinfo{volume}{D77}},
  \bibinfo{pages}{081501} (\bibinfo{year}{2008}), \eprint{0712.1716}.

\bibitem[{\citenamefont{Damour et~al.}(2000)\citenamefont{Damour, Jaranowski,
  and Sch{\"a}fer}}]{Damour:2000we}
\bibinfo{author}{\bibfnamefont{T.}~\bibnamefont{Damour}},
  \bibinfo{author}{\bibfnamefont{P.}~\bibnamefont{Jaranowski}},
  \bibnamefont{and}
  \bibinfo{author}{\bibfnamefont{G.}~\bibnamefont{Sch{\"a}fer}},
  \bibinfo{journal}{Phys. Rev.} \textbf{\bibinfo{volume}{D62}},
  \bibinfo{pages}{084011} (\bibinfo{year}{2000}), \eprint{gr-qc/0005034}.

\bibitem[{\citenamefont{Barausse et~al.}(2012)\citenamefont{Barausse, Buonanno,
  and Le~Tiec}}]{Barausse:2011dq}
\bibinfo{author}{\bibfnamefont{E.}~\bibnamefont{Barausse}},
  \bibinfo{author}{\bibfnamefont{A.}~\bibnamefont{Buonanno}}, \bibnamefont{and}
  \bibinfo{author}{\bibfnamefont{A.}~\bibnamefont{Le~Tiec}},
  \bibinfo{journal}{Phys.Rev.} \textbf{\bibinfo{volume}{D85}},
  \bibinfo{pages}{064010} (\bibinfo{year}{2012}), \eprint{1111.5610}.

\bibitem[{\citenamefont{Damour et~al.}(2013)\citenamefont{Damour, Nagar, and
  Bernuzzi}}]{Damour:2012ky}
\bibinfo{author}{\bibfnamefont{T.}~\bibnamefont{Damour}},
  \bibinfo{author}{\bibfnamefont{A.}~\bibnamefont{Nagar}}, \bibnamefont{and}
  \bibinfo{author}{\bibfnamefont{S.}~\bibnamefont{Bernuzzi}},
  \bibinfo{journal}{Phys.Rev.} \textbf{\bibinfo{volume}{D87}},
  \bibinfo{pages}{084035} (\bibinfo{year}{2013}), \eprint{1212.4357}.

\bibitem[{\citenamefont{{Bini} and {Damour}}(2013)}]{Bini:2013PhRvD..87l1501B}
\bibinfo{author}{\bibfnamefont{D.}~\bibnamefont{{Bini}}} \bibnamefont{and}
  \bibinfo{author}{\bibfnamefont{T.}~\bibnamefont{{Damour}}},
  \bibinfo{journal}{\prd} \textbf{\bibinfo{volume}{87}}, \bibinfo{eid}{121501}
  (\bibinfo{year}{2013}), \eprint{1305.4884}.

\bibitem[{\citenamefont{Pan et~al.}(2014)\citenamefont{Pan, Buonanno,
  Taracchini, Kidder, Mroue et~al.}}]{Pan:2013rra}
\bibinfo{author}{\bibfnamefont{Y.}~\bibnamefont{Pan}},
  \bibinfo{author}{\bibfnamefont{A.}~\bibnamefont{Buonanno}},
  \bibinfo{author}{\bibfnamefont{A.}~\bibnamefont{Taracchini}},
  \bibinfo{author}{\bibfnamefont{L.~E.} \bibnamefont{Kidder}},
  \bibinfo{author}{\bibfnamefont{A.~H.} \bibnamefont{Mroue}},
  \bibnamefont{et~al.}, \bibinfo{journal}{Phys.Rev.}
  \textbf{\bibinfo{volume}{D89}}, \bibinfo{pages}{084006}
  (\bibinfo{year}{2014}), \eprint{1307.6232}.

\bibitem[{\citenamefont{Damour et~al.}(2008{\natexlab{b}})\citenamefont{Damour,
  Jaranowski, and Sch{\"a}fer}}]{Damour:2008qf}
\bibinfo{author}{\bibfnamefont{T.}~\bibnamefont{Damour}},
  \bibinfo{author}{\bibfnamefont{P.}~\bibnamefont{Jaranowski}},
  \bibnamefont{and}
  \bibinfo{author}{\bibfnamefont{G.}~\bibnamefont{Sch{\"a}fer}},
  \bibinfo{journal}{Phys.Rev.} \textbf{\bibinfo{volume}{D78}},
  \bibinfo{pages}{024009} (\bibinfo{year}{2008}{\natexlab{b}}),
  \eprint{0803.0915}.

\bibitem[{\citenamefont{{Balmelli} and
  {Jetzer}}(2013)}]{Balmelli:2013PhRvD..87l4036B}
\bibinfo{author}{\bibfnamefont{S.}~\bibnamefont{{Balmelli}}} \bibnamefont{and}
  \bibinfo{author}{\bibfnamefont{P.}~\bibnamefont{{Jetzer}}},
  \bibinfo{journal}{\prd} \textbf{\bibinfo{volume}{87}}, \bibinfo{eid}{124036}
  (\bibinfo{year}{2013}), \eprint{1305.5674}.

\bibitem[{\citenamefont{Paschalidis et~al.}(2012)\citenamefont{Paschalidis,
  Etienne, and Shapiro}}]{Paschalidis:2012ff}
\bibinfo{author}{\bibfnamefont{V.}~\bibnamefont{Paschalidis}},
  \bibinfo{author}{\bibfnamefont{Z.~B.} \bibnamefont{Etienne}},
  \bibnamefont{and} \bibinfo{author}{\bibfnamefont{S.~L.}
  \bibnamefont{Shapiro}}, \bibinfo{journal}{Phys.Rev.}
  \textbf{\bibinfo{volume}{D86}}, \bibinfo{pages}{064032}
  (\bibinfo{year}{2012}), \eprint{1208.5487}.

\bibitem[{\citenamefont{{Hotokezaka} et~al.}(2013)\citenamefont{{Hotokezaka},
  {Kiuchi}, {Kyutoku}, {Muranushi}, {Sekiguchi}, {Shibata}, and
  {Taniguchi}}}]{Hotokezaka:2013PhRvD..88d4026H}
\bibinfo{author}{\bibfnamefont{K.}~\bibnamefont{{Hotokezaka}}},
  \bibinfo{author}{\bibfnamefont{K.}~\bibnamefont{{Kiuchi}}},
  \bibinfo{author}{\bibfnamefont{K.}~\bibnamefont{{Kyutoku}}},
  \bibinfo{author}{\bibfnamefont{T.}~\bibnamefont{{Muranushi}}},
  \bibinfo{author}{\bibfnamefont{Y.-i.} \bibnamefont{{Sekiguchi}}},
  \bibinfo{author}{\bibfnamefont{M.}~\bibnamefont{{Shibata}}},
  \bibnamefont{and}
  \bibinfo{author}{\bibfnamefont{K.}~\bibnamefont{{Taniguchi}}},
  \bibinfo{journal}{\prd} \textbf{\bibinfo{volume}{88}}, \bibinfo{eid}{044026}
  (\bibinfo{year}{2013}), \eprint{1307.5888}.

\bibitem[{\citenamefont{Deaton et~al.}(2013)\citenamefont{Deaton, Duez,
  Foucart, O'Connor, Ott et~al.}}]{Deaton:2013sla}
\bibinfo{author}{\bibfnamefont{M.~B.} \bibnamefont{Deaton}},
  \bibinfo{author}{\bibfnamefont{M.~D.} \bibnamefont{Duez}},
  \bibinfo{author}{\bibfnamefont{F.}~\bibnamefont{Foucart}},
  \bibinfo{author}{\bibfnamefont{E.}~\bibnamefont{O'Connor}},
  \bibinfo{author}{\bibfnamefont{C.~D.} \bibnamefont{Ott}},
  \bibnamefont{et~al.}, \bibinfo{journal}{Astrophys.J.}
  \textbf{\bibinfo{volume}{776}}, \bibinfo{pages}{47} (\bibinfo{year}{2013}),
  \eprint{1304.3384}.

\bibitem[{\citenamefont{Stergioulas et~al.}(2011)\citenamefont{Stergioulas,
  Bauswein, Zagkouris, and Janka}}]{Stergioulas:2011gd}
\bibinfo{author}{\bibfnamefont{N.}~\bibnamefont{Stergioulas}},
  \bibinfo{author}{\bibfnamefont{A.}~\bibnamefont{Bauswein}},
  \bibinfo{author}{\bibfnamefont{K.}~\bibnamefont{Zagkouris}},
  \bibnamefont{and} \bibinfo{author}{\bibfnamefont{H.-T.} \bibnamefont{Janka}}
  (\bibinfo{year}{2011}), \eprint{1105.0368}.

\bibitem[{\citenamefont{Baiotti
  et~al.}(2009{\natexlab{a}})\citenamefont{Baiotti, Bernuzzi, Corvino,
  De~Pietri, and Nagar}}]{Baiotti:2008nf}
\bibinfo{author}{\bibfnamefont{L.}~\bibnamefont{Baiotti}},
  \bibinfo{author}{\bibfnamefont{S.}~\bibnamefont{Bernuzzi}},
  \bibinfo{author}{\bibfnamefont{G.}~\bibnamefont{Corvino}},
  \bibinfo{author}{\bibfnamefont{R.}~\bibnamefont{De~Pietri}},
  \bibnamefont{and} \bibinfo{author}{\bibfnamefont{A.}~\bibnamefont{Nagar}},
  \bibinfo{journal}{Phys. Rev.} \textbf{\bibinfo{volume}{D79}},
  \bibinfo{pages}{024002} (\bibinfo{year}{2009}{\natexlab{a}}),
  \eprint{0808.4002}.

\bibitem[{\citenamefont{Dimmelmeier et~al.}(2006)\citenamefont{Dimmelmeier,
  Stergioulas, and Font}}]{Dimmelmeier:2005zk}
\bibinfo{author}{\bibfnamefont{H.}~\bibnamefont{Dimmelmeier}},
  \bibinfo{author}{\bibfnamefont{N.}~\bibnamefont{Stergioulas}},
  \bibnamefont{and} \bibinfo{author}{\bibfnamefont{J.~A.} \bibnamefont{Font}},
  \bibinfo{journal}{Mon. Not. Roy. Astron. Soc.}
  \textbf{\bibinfo{volume}{368}}, \bibinfo{pages}{1609} (\bibinfo{year}{2006}),
  \eprint{astro-ph/0511394}.

\bibitem[{\citenamefont{Bauswein and Janka}(2012)}]{Bauswein:2011tp}
\bibinfo{author}{\bibfnamefont{A.}~\bibnamefont{Bauswein}} \bibnamefont{and}
  \bibinfo{author}{\bibfnamefont{H.-T.} \bibnamefont{Janka}},
  \bibinfo{journal}{Phys.Rev.Lett.} \textbf{\bibinfo{volume}{108}},
  \bibinfo{pages}{011101} (\bibinfo{year}{2012}), \eprint{1106.1616}.

\bibitem[{\citenamefont{Foucart et~al.}(2013)\citenamefont{Foucart, Buchman,
  Duez, Grudich, Kidder et~al.}}]{Foucart:2013psa}
\bibinfo{author}{\bibfnamefont{F.}~\bibnamefont{Foucart}},
  \bibinfo{author}{\bibfnamefont{L.}~\bibnamefont{Buchman}},
  \bibinfo{author}{\bibfnamefont{M.~D.} \bibnamefont{Duez}},
  \bibinfo{author}{\bibfnamefont{M.}~\bibnamefont{Grudich}},
  \bibinfo{author}{\bibfnamefont{L.~E.} \bibnamefont{Kidder}},
  \bibnamefont{et~al.}, \bibinfo{journal}{Phys.Rev.}
  \textbf{\bibinfo{volume}{D88}}, \bibinfo{pages}{064017}
  (\bibinfo{year}{2013}), \eprint{1307.7685}.

\bibitem[{\citenamefont{Bernuzzi et~al.}(2014)\citenamefont{Bernuzzi, Nagar,
  Balmelli, Dietrich, and Ujevic}}]{Bernuzzi:2014kca}
\bibinfo{author}{\bibfnamefont{S.}~\bibnamefont{Bernuzzi}},
  \bibinfo{author}{\bibfnamefont{A.}~\bibnamefont{Nagar}},
  \bibinfo{author}{\bibfnamefont{S.}~\bibnamefont{Balmelli}},
  \bibinfo{author}{\bibfnamefont{T.}~\bibnamefont{Dietrich}}, \bibnamefont{and}
  \bibinfo{author}{\bibfnamefont{M.}~\bibnamefont{Ujevic}},
  \bibinfo{journal}{Phys.Rev.Lett.} \textbf{\bibinfo{volume}{112}},
  \bibinfo{pages}{201101} (\bibinfo{year}{2014}), \eprint{1402.6244}.

\bibitem[{\citenamefont{Ajith et~al.}(2011)\citenamefont{Ajith, Hannam, Husa,
  Chen, Br{\"u}gmann et~al.}}]{Ajith:2009bn}
\bibinfo{author}{\bibfnamefont{P.}~\bibnamefont{Ajith}},
  \bibinfo{author}{\bibfnamefont{M.}~\bibnamefont{Hannam}},
  \bibinfo{author}{\bibfnamefont{S.}~\bibnamefont{Husa}},
  \bibinfo{author}{\bibfnamefont{Y.}~\bibnamefont{Chen}},
  \bibinfo{author}{\bibfnamefont{B.}~\bibnamefont{Br{\"u}gmann}},
  \bibnamefont{et~al.}, \bibinfo{journal}{Phys.Rev.Lett.}
  \textbf{\bibinfo{volume}{106}}, \bibinfo{pages}{241101}
  (\bibinfo{year}{2011}), \eprint{0909.2867}.

\bibitem[{\citenamefont{Hannam et~al.}(2010)\citenamefont{Hannam, Husa, Ohme,
  M{\"u}ller, and Br{\"u}gmann}}]{Hannam:2010ec}
\bibinfo{author}{\bibfnamefont{M.}~\bibnamefont{Hannam}},
  \bibinfo{author}{\bibfnamefont{S.}~\bibnamefont{Husa}},
  \bibinfo{author}{\bibfnamefont{F.}~\bibnamefont{Ohme}},
  \bibinfo{author}{\bibfnamefont{D.}~\bibnamefont{M{\"u}ller}},
  \bibnamefont{and}
  \bibinfo{author}{\bibfnamefont{B.}~\bibnamefont{Br{\"u}gmann}},
  \bibinfo{journal}{Phys. Rev.} \textbf{\bibinfo{volume}{D82}},
  \bibinfo{pages}{124008} (\bibinfo{year}{2010}), \eprint{1007.4789}.

\bibitem[{\citenamefont{Santamaria et~al.}(2010)\citenamefont{Santamaria, Ohme,
  Ajith, Br{\"u}gmann, Dorband et~al.}}]{Santamaria:2010yb}
\bibinfo{author}{\bibfnamefont{L.}~\bibnamefont{Santamaria}},
  \bibinfo{author}{\bibfnamefont{F.}~\bibnamefont{Ohme}},
  \bibinfo{author}{\bibfnamefont{P.}~\bibnamefont{Ajith}},
  \bibinfo{author}{\bibfnamefont{B.}~\bibnamefont{Br{\"u}gmann}},
  \bibinfo{author}{\bibfnamefont{N.}~\bibnamefont{Dorband}},
  \bibnamefont{et~al.}, \bibinfo{journal}{Phys.Rev.}
  \textbf{\bibinfo{volume}{D82}}, \bibinfo{pages}{064016}
  (\bibinfo{year}{2010}), \eprint{1005.3306}.

\bibitem[{\citenamefont{Hannam et~al.}(2008)\citenamefont{Hannam, Husa,
  Br{\"u}gmann, and Gopakumar}}]{Hannam:2007wf}
\bibinfo{author}{\bibfnamefont{M.}~\bibnamefont{Hannam}},
  \bibinfo{author}{\bibfnamefont{S.}~\bibnamefont{Husa}},
  \bibinfo{author}{\bibfnamefont{B.}~\bibnamefont{Br{\"u}gmann}},
  \bibnamefont{and}
  \bibinfo{author}{\bibfnamefont{A.}~\bibnamefont{Gopakumar}},
  \bibinfo{journal}{Phys. Rev.} \textbf{\bibinfo{volume}{D78}},
  \bibinfo{pages}{104007} (\bibinfo{year}{2008}), \eprint{0712.3787}.

\bibitem[{\citenamefont{Favata}(2014)}]{Favata:2013rwa}
\bibinfo{author}{\bibfnamefont{M.}~\bibnamefont{Favata}},
  \bibinfo{journal}{Phys.Rev.Lett.} \textbf{\bibinfo{volume}{112}},
  \bibinfo{pages}{101101} (\bibinfo{year}{2014}), \eprint{1310.8288}.

\bibitem[{\citenamefont{Baiotti
  et~al.}(2009{\natexlab{b}})\citenamefont{Baiotti, Giacomazzo, and
  Rezzolla}}]{Baiotti:2009gk}
\bibinfo{author}{\bibfnamefont{L.}~\bibnamefont{Baiotti}},
  \bibinfo{author}{\bibfnamefont{B.}~\bibnamefont{Giacomazzo}},
  \bibnamefont{and} \bibinfo{author}{\bibfnamefont{L.}~\bibnamefont{Rezzolla}},
  \bibinfo{journal}{Class.Quant.Grav.} \textbf{\bibinfo{volume}{26}},
  \bibinfo{pages}{114005} (\bibinfo{year}{2009}{\natexlab{b}}),
  \eprint{0901.4955}.

\bibitem[{\citenamefont{{Eric Gourgoulhon, Philippe Grandcl\'{e}ment,
  Jean-Alain Marck, J\'{e}r\^{o}me Novak and Keisuke Taniguchi}}()}]{LORENE}
\bibinfo{author}{\bibnamefont{{Eric Gourgoulhon, Philippe Grandcl\'{e}ment,
  Jean-Alain Marck, J\'{e}r\^{o}me Novak and Keisuke Taniguchi}}},
  \bibinfo{note}{{Paris Observatory, Meudon section - LUTH laboratory}},
  \urlprefix\url{{http://www.lorene.obspm.fr/}}.

\bibitem[{\citenamefont{{Berger} and {Colella}}(1989)}]{Berger:1989}
\bibinfo{author}{\bibfnamefont{M.~J.} \bibnamefont{{Berger}}} \bibnamefont{and}
  \bibinfo{author}{\bibfnamefont{P.}~\bibnamefont{{Colella}}},
  \bibinfo{journal}{Journal of Computational Physics}
  \textbf{\bibinfo{volume}{82}}, \bibinfo{pages}{64} (\bibinfo{year}{1989}).

\bibitem[{\citenamefont{East et~al.}(2012)\citenamefont{East, Pretorius, and
  Stephens}}]{East:2011aa}
\bibinfo{author}{\bibfnamefont{W.~E.} \bibnamefont{East}},
  \bibinfo{author}{\bibfnamefont{F.}~\bibnamefont{Pretorius}},
  \bibnamefont{and} \bibinfo{author}{\bibfnamefont{B.~C.}
  \bibnamefont{Stephens}}, \bibinfo{journal}{Phys.Rev.}
  \textbf{\bibinfo{volume}{D85}}, \bibinfo{pages}{124010}
  (\bibinfo{year}{2012}), \eprint{1112.3094}.

\bibitem[{\citenamefont{Reisswig et~al.}(2013)\citenamefont{Reisswig, Haas,
  Ott, Abdikamalov, M{\"o}sta et~al.}}]{Reisswig:2012nc}
\bibinfo{author}{\bibfnamefont{C.}~\bibnamefont{Reisswig}},
  \bibinfo{author}{\bibfnamefont{R.}~\bibnamefont{Haas}},
  \bibinfo{author}{\bibfnamefont{C.}~\bibnamefont{Ott}},
  \bibinfo{author}{\bibfnamefont{E.}~\bibnamefont{Abdikamalov}},
  \bibinfo{author}{\bibfnamefont{P.}~\bibnamefont{M{\"o}sta}},
  \bibnamefont{et~al.}, \bibinfo{journal}{Phys.Rev.}
  \textbf{\bibinfo{volume}{D87}}, \bibinfo{pages}{064023}
  (\bibinfo{year}{2013}), \eprint{1212.1191}.

\end{thebibliography}
\end{document}